\documentclass[11pt,a4paper]{article}

\usepackage[english]{babel}
\usepackage[T1]{fontenc}
\usepackage[utf8]{inputenc}

\usepackage{lmodern}
\usepackage{microtype}

\usepackage{amsmath,amssymb,amsthm}
\usepackage{graphicx}
\usepackage{natbib}

\usepackage{url}

\usepackage[a4paper,margin=2.5cm]{geometry}
\title{Predicting disease severity and large-scale spread from coupled severity measurements and imperfect indicators: Application to beet yellows}

\author{
Baptiste Oger\thanks{Corresponding author: baptiste.oger@inrae.fr}\\
INRAE, BioSP, Avignon, France
\and
César Martinez\\
INRAE, Ecodev, Avignon, France
\and
François Joudelat\\
Institut Technique de la Betterave, Paris, France
\and
Samuel Soubeyrand\thanks{These authors contributed equally as senior authors.}\\
INRAE, BioSP, Avignon, France
\and
Lionel Benoit\footnotemark[2]\\
INRAE, BioSP, Avignon, France
}

\date{}

\begin{document}

\maketitle

\begin{abstract}
Whether in human, animal, or plant health, effective disease management requires the ability to characterize disease dynamics across space and time. In this context, integrating indirect indicators with broad spatio-temporal coverage, even when they are noisy, can provide valuable complementary information to direct measurements, which are often sparse because they are more costly or intrusive to collect. In this article, we propose a statistical framework to leverage such indirect indicators to predict disease severity at the individual or local-scale level and reconstruct large-scale disease dynamics. This two-step approach is able to account for the specific characteristics of disease severity observations, including zero inflation and spatio-temporal structure. The first step relies on a stacked hurdle model based on multiple random forests to locally predict disease severity from the available indirect indicators. In the second step a semi-parametric spatio-temporal model is used to reconstruct large-scale epidemiological dynamics over space and time from the indicators-based predictions. The proposed methodology is designed to be both generic and modular, and is illustrated by a case study in plant health. This case study focuses on the monitoring of sugar beet yellows disease in France between 2019 and 2023 by combining sparse field measurements and satellite-based remote sensing data.
\end{abstract}

\vspace{0.5em}


\maketitle

\section{Introduction}

Monitoring disease severity and spread across space and time is a central challenge in epidemiology, spanning human, animal, and plant health \citep{robertson_review_2010, moustakas_spatio-temporal_2017, cunniffe_thirteen_2015}. A key challenge in pursuing this objective is that direct measurements of severity such as clinical scores, pathogen loads, tissue damage assessments, symptom coverages, or local prevalence, are often costly, invasive, or impractical to collect at scale. Hence, they are often available only for a limited subset of individuals or locations due to logistical, economic, or technical constraints.

Large-scale surveillance systems may therefore benefit from indirect indicators of disease, such as imperfect diagnostic tests, non-specific symptoms, proxy measures of damage, or remotely sensed observations \citep{getsova_syndromic_2025, berezowski_complex_2019, oerke_remote_2020}. These indicators are nevertheless usually heterogeneous, noisy, and only indirectly related to the latent, quantitative disease process. A central inferential challenge is consequently to use such indicators to estimate disease severity at the individual or local scale, and to propagate this information to infer disease spread over larger spatial domains.

Such an inference task raises methodological challenges stemming from four key features of surveillance data. First, disease severity is often zero-inflated, with a large proportion of zero observations corresponding to disease absence, the remaining observations corresponding to positive values of severity \citep{soogun_spatiotemporal_2022, bradhurst_hybrid_2015, bock_plant_2022, barreto_disease_2023}. Statistical models must therefore account for zero inflation while preserving sensitivity to non-zero severity levels. Second, indirect indicators are imperfect and may be biased, noisy, or affected by unobserved confounders related to host, environment, or management context. Third, disease dynamics are structured in space and time, requiring models that borrow strength across neighbouring units to produce coherent local estimates and capture large-scale spread. Fourth, logistic considerations require practitioners to design a cost-effective balance between accurate field disease severity measurements and indirect indicator observations when estimating disease spread.

Existing modelling approaches address some of these challenges. Zero-inflated and hurdle models provide flexible tools to account for excess zeros in disease severity data \citep{liu_statistical_2019}, and can be embedded within hierarchical modelling frameworks to represent underlying epidemiological processes while accommodating zero inflation \citep{arab_spatial_2015, wang_bayesian_2015}. Spatio-temporal statistical models capture the spatial spread of epidemics across landscapes \citep{lawson_statistical_2013}. In parallel, machine learning approaches have proven effective for predicting disease presence or severity from heterogeneous and imperfect predictors \citep{bellinger_systematic_2017}. However, no approach jointly addresses these four challenges within a unified framework while remaining flexible enough to accommodate diverse epidemiological settings. In this paper, we integrate existing statistical methods to build such a framework in a two-step approach: in step 1 a stacked hurdle model \citep{feng_comparison_2021, pavlyshenko_using_2018} grounded on multiple random forests \citep{tong_spectralspatial_2022} links individual-level severity measurements with indirect indicators including spatially informed meta-features; in step 2 a spatio-temporal semi-parametric model combining a logistic growth model \citep{werker_modelling_1998} with a Nadaraya-Watson smoother \citep{lawson_statistical_2013} maps the epidemics at large scale by leveraging individual-level estimations obtained in step 1. This framework is designed to be general and ensure broad applicability so it can be transferred across regions, time, diseases, and data sources, and can be adapted to varying epidemiological contexts. Its modularity also facilitates the integration of additional data and alternative model components as needed, while maintaining a balance between predictive performance and interpretability.

In what follows, Section \ref{sec:pb} describes the case study that motivated the design of the modelling framework at the core of the present work. It deals with sugar beet yellows, a plant disease for which we propose to complement the field monitoring by satellite imagery. Section \ref{sec:models} describes the modelling framework and its implementation. Section \ref{sec:results} presents the results and demonstrates the added value of satellite imagery. Finally, section \ref{sec:discu} discusses the results, the advantages and limitations of our approach, and replaces it in a broader epidemiological context.

\section{Motivating problem}
\label{sec:pb}

Sugar beet is the second most important crop for sugar production in the world and beet yellows disease represents the main threat to sugar beet production. On affected plant, this disease, results in stunted growth and reduced sugar content. In infected fields, yield losses can reach 20\% to 40\%, depending on the viral strain and the phenological stage at which the plants are affected \citep{hossain_new_2021}. Although chemical seed treatments with neonicotinoids substantially reduced the impact of beet yellows for several decades, the ban imposed on these products in Europe in 2018 resulted with a resurgence of sugar beet yellows and significant yield losses in several regions \citep{verheggen_producing_2022}. In this context, improving the capacity to monitor beet yellows at large spatial scales and throughout the entire growing season has become a priority for farmers, advisers, and plant-health services. Early detection of disease severity enable timely prophylactic interventions, to prevent disease propagation and worsening, and provide objective information on potential yield-loss.

Traditional methods for monitoring and assessing the severity of beet yellows largely rely on field observations carried out by farmers or other stakeholders, as well as on sample collection and laboratory testing when viral strain identification is required. These methods remain limited and too costly to enable precise tracking of the development and spread of the disease across space and time at the regional to country scale. At the same time, the expansion of remote sensing applications in agriculture opens new perspectives for epidemiological modelling by enabling the monitoring of beet yellows dynamics across space and time at large spatial scales \citep{mikaberidze_opportunities_2025}. Recent studies have demonstrated the potential of these data for monitoring other sugar beet pests \citep{hillnhutter_remote_2011} as well as for detecting virus-induced yellowing in other crops \citep{guo_recognition_2022, zibrat_detection_2024}. However, no remote-sensing-based framework exists yet for detecting or monitoring beet yellows specifically. More generally, there is a growing need for statistical frameworks capable of exploiting multispectral time-series data to track foliar disease dynamics at both field and landscape scales. Satellite-derived indicators provide only indirect and imperfect information on disease severity, requiring statistical models able to account for key epidemiological data characteristics, such as zero inflation and spatio-temporal structure, while remaining robust to noisy and incomplete inputs. More information on the motivating problem are available in Appendix A.

\subsection{Data}

\paragraph{Field observations.}

\begin{figure} 
\centering
\includegraphics[width=17cm,trim={3cm 0 2cm 0},clip]{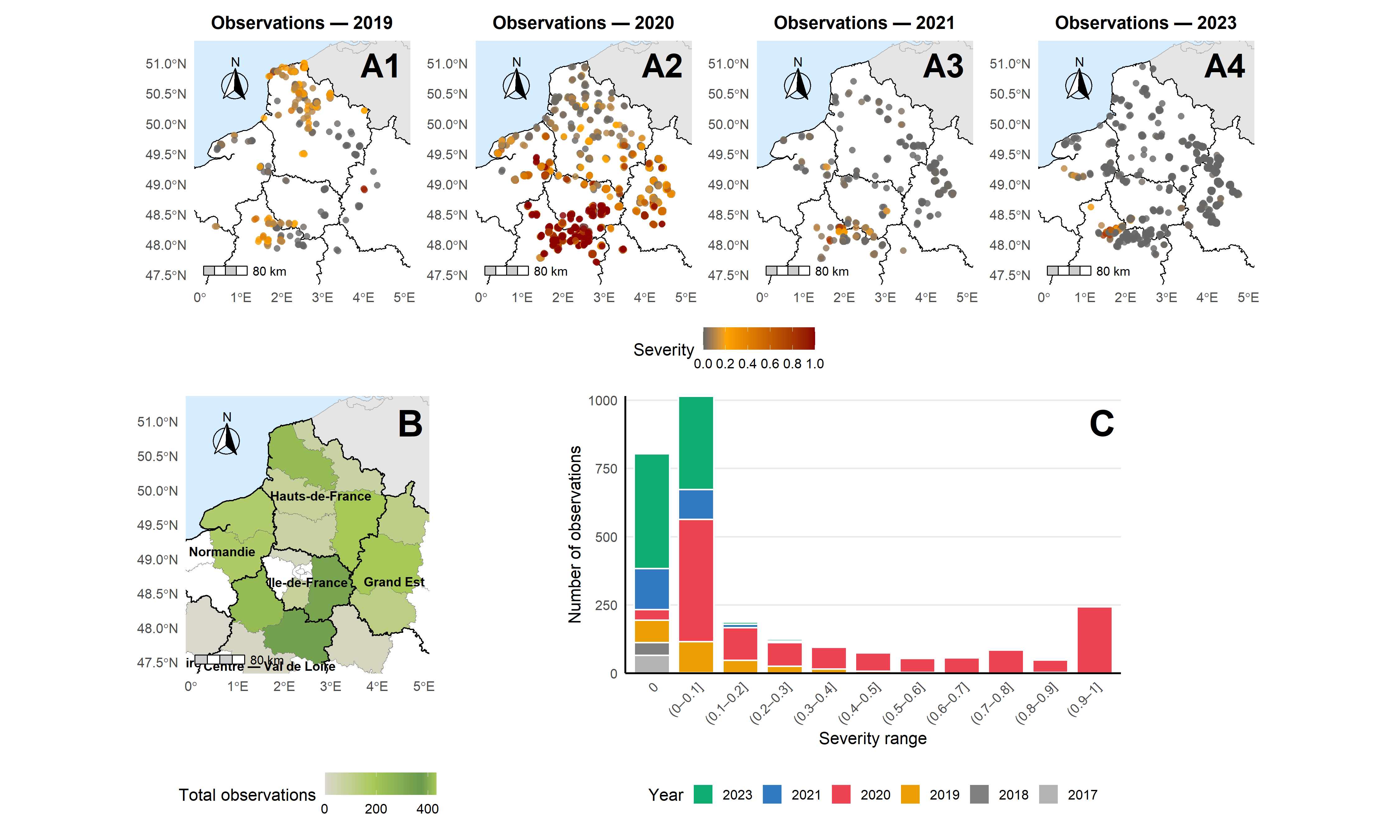}
\caption{\label{fig:illus_data} A1 to A4: Spatial distribution of observations of sugar beet yellows severity for available years between 2019 and 2023; B: Main regions with total number of observations per department and; C: distribution of severity values.}
\end{figure}

Beet yellows severity data were collected through the French sugar beet epidemiosurveillance network (VigiBet®), coordinated by the French Technical Institute of Sugar Beet (ITB : https://www.itbfr.org/). Observations were carried out in France between 2017 and 2023, across the main sugar beet production areas: Hauts-de-France, Grand-Est, Normandie, Île-de-France, and Centre-Val-de-Loire regions (Figure \ref{fig:illus_data}). These observations were carried out by technicians from various organisations, by farmers or by students participating in monitoring activities. The 2017 and 2018 datasets are more limited, consisting only of fields without disease symptoms, as these years correspond to a period when neonicotinoid seed treatments were still authorised and widely used in France. Data from 2022 were excluded as some of the covariates, available for the other years, could not be retrieved for that year.

Each record includes a severity observation, the date of observation, spatial coordinates, and the name of the cultivated beet variety. The severity value represents a visual assessment of the proportion of the field area affected by beet yellows disease, ranging from 0 to 1. A value of 0 indicates no visible symptoms and a value of 1 corresponds to a situation in which the entire field is affected. It should be noted that, due to the visual nature of the assessments and the diversity of observers involved, some unquantified uncertainty may be present in the severity measurements. The complete dataset comprises 2,835 observations from 778 fields, some of which were surveyed multiple times within the same season to capture temporal disease dynamics.

\paragraph{Satellite data.}

For each field with at least one in-situ observation, its detailed shape was manually drawn on QGIS software (QGIS Association, http://www.qgis.org) based on the spatial coordinates of the observations, the French Land Parcel Identification System (LPIS, Registre Parcellaire Graphique - https://geoservices.ign.fr/rpg) and the THEIA Sentinel-2 cloudless mosaics from the THEIA platform (https://www.theia-land.fr). Its field geometry was attributed a unique identifier matching with field observations. Sentinel-2 data where then retrieved for each field from the THEIA platform, which was chosen for the quality of its level-2 radiometric correction \citep{hagolle_multi-temporal_2010}. For each severity observation, the Sentinel-2 image, corresponding to the closest acquisition date with no cloud or cloud-shadow contamination over the field, was retained. The ten Sentinel-2 reflectance bands available at 10 m and 20 m spatial resolution were then clipped to the beet field geometries, without applying any buffer. The bands originally at 20 m were resampled to 10 m before computing the indices. Finally, 29 spectral indices described in Appendix B were computed for all pixels and averaged at the field-level.

\paragraph{Final dataset.}

The final dataset associates each field observation of beet yellows severity with 50 covariates: 10 raw reflectance values; 29 indices derived from these reflectances; 4 covariates relative to Sentinel-2 metadata (two viewing angles, the image acquisition date and the number of day between field observation and satellite image); and 7 covariates relative to field and observation metadata (geographic coordinates, beet variety, observation date, day of the year, month and year). Final list of covariates is available in Appendix B.

\section{modelling framework}
\label{sec:models}

This section details the modelling framework. Implementation choices related to the case study are provided in the Appendix C. The framework is composed of two steps. In a first step, a stacked hurdle model based on Random Forest learners is used to predict local disease severity from indirect indicators. In a second step, a semi-parametric spatio-temporal model combining an interpretable temporal structure with spatial smoothing propagates these local predictions to reconstruct epidemic dynamics at larger spatio-temporal scales.

\subsection{A stacked hurdle model with spatial meta-features for local-scale disease prediction}

\begin{figure} 
\centering
\includegraphics[width=15cm]{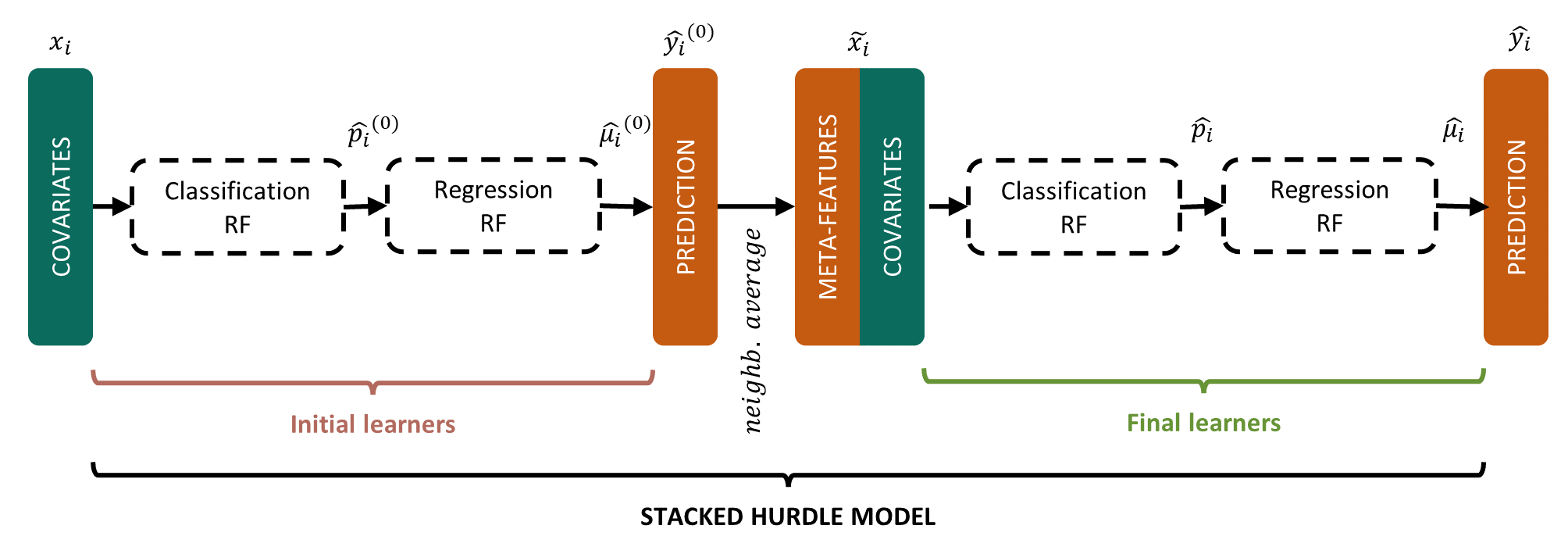}
\caption{\label{fig:Method_FieldScale} Global workflow for the stacked hurdle model.}
\end{figure}

To model disease severity at the local scale from covariates, we adopt a \emph{stacked regression} approach: an initial learner produces predictions that are subsequently transformed into meta-features and supplied to the final learner; see Figure \ref{fig:Method_FieldScale}. An hurdle factorization is used at both levels for handling zero-inflation in observed disease severity \citep{feng_comparison_2021}. The meta-features spatially aggregate predictions at multiple spatial scales, which allows the model to consider a neighbourhood context without an explicit model of spatial covariance. Here random forests are used to perform non-linear regressions, but the stacking and hurdle frameworks are model-agnostic and could rely on alternative learners.

\paragraph{Notation.}
Let $\{(x_i, s_i, t_i, y_i)\}_{i=1}^n$ denote observations, where
$x_i \in \mathbb{R}^q$ ($q\in{\mathbb{N}}^*$) is the vector of covariates for observation $i$, $s_i$ is its spatial location, $t_i$ is
its time, and $y_i \in [0,1]$ is the local-scale disease severity. We use capital letters (e.g., $Y_i$) to denote random variables and lowercase to denote the related observations (e.g., $y_i$). 
Spatial location $s_i$ and time $t_i$ are used to characterize spatial and temporal relationships between observations. Their numerical representations can also be included among the covariates in $x_i$ (as this is the case in the motivating problem; see Section \ref{sec:pb}). We also define the binary variable $Z_i = \mathbf{1}(Y_i>0)$.

\paragraph{Initial learners: hurdle Random Forest.}
Initially, a two-component (hurdle) model is fitted:
\begin{enumerate}
  \item A classification learner for the occurrence of the disease:
\begin{equation}
    \label{eq:classif_initial_learner}
    \hat p_i^{(0)} = \hat {\mathbb{P}}(Z_i=1 \mid x_i).
\end{equation}
  
  \item A regression learner for the severity conditional to the presence of the disease:
  \begin{equation}
    \label{eq:reg_initial_learner}
    \hat \mu_i^{(0)} = \widehat{\mathbb{E}}(Y_i \mid Z_i=1, x_i),
    \quad\text{fitted on }\{i\in\{1,\ldots,n\}: y_i>0\}.
\end{equation}

\end{enumerate}
The initial point prediction of severity is the product
$\hat y_i^{(0)} = \hat p_i^{(0)} \,\hat \mu_i^{(0)}$, which estimates
$\mathbb{E}(Y_i \mid x_i)$ under the hurdle factorization \citep{liu_statistical_2019, rozanec_dealing_2025}.
Both learners are Random Forests \citep{francisco_hybrid_2024} with hyperparameters selected by cross-validation \citep{hastie_elements_2009}. Cross validation is detailed in the implementation section (see Appendix C). A grid of candidate values for the number of variables randomly sampled at each split ($mtry$) and the number of trees ($ntrees$) was defined and evaluated separately for each learner.

For the classification learner, hyperparameter combinations were assessed using the overall accuracy (OA): 
\begin{equation}
\label{eq:oa}
\mathrm{OA} = \frac{1}{n} \sum_{i=1}^{n} \mathbf{1}(z_i = \hat{z}_i^{(0)}), \qquad \text{where } \hat{z}_i^{(0)}=\mathbf{1}(\hat{p}_i^{(0)} > 0.5) .
\end{equation}

For the regression learner, the best hyperparameters were selected using the root mean squared error (RMSE) as the performance criterion:
\begin{equation}
\label{eq:rmse}
\mathrm{RMSE} = \sqrt{\frac{1}{n} \sum_{i=1}^{n} (y_i - \hat{y}_i^{(0)})^2} .
\end{equation}
The optimal parameters ($mtry$, $ntrees$) maximizing OA (resp. minimizing RMSE) were used to fit the classification learner (resp. regression learner).

\paragraph{Final learners: stacked hurdle Random Forest.}

Including covariates derived from neighbouring observations has proven to be an effective way to capture the local spatial dependence in predictive models \citep{ghimire_contextual_2010, hengl_random_2018, mariano_random_2021}. Here, we define the \emph{meta-features} as neighbourhood averages of the initial predictions for a neighbourhood radius $d\ge 0$:
\begin{equation}
    \label{eq:meta_feature}
    \tilde y_{id} = \frac{1}{|\mathcal{N}_{id}|}\sum_{j\in\mathcal{N}_{id}} \hat y_j^{(0)},
\end{equation}
with the same-year neighbour set
$\mathcal{N}_{id} =\{j\in\{1,\ldots,n\}:j\neq i, \ \mathrm{dist}(s_i,s_j)\le d,\ \mathrm{year}(t_j)=\mathrm{year}(t_i)\}$. 

Let \(0 = d_1 < \cdots < d_L\) denote a set of \(L\) neighbourhood radii used to
summarise model predictions at multiple spatial scales. The \emph{augmented} covariate vector, made of initial covariates and the stacked meta-features, is defined as:
\[
  \tilde x_i = \big(x_i^\top, \tilde y_{id_1},\ldots, \tilde y_{id_L})^\top \in \mathbb{R}^{q+L}.
\]

 We refit a two-component (hurdle) model, using $\tilde x_i$ instead of $x_i$:
\begin{enumerate}
  \item A new classification learner for the occurrence:
  \begin{equation}
    \label{eq:classif_final_learner}
    \hat p_i = \hat {\mathbb{P}}_0(Z_i=1 \mid \tilde x_i).
\end{equation}
  \item A new regression learner for the positive part:
    \begin{equation}
    \label{eq:reg_final_learner}
    \hat \mu_i = \widehat{\mathbb{E}}_0(Y_i \mid Z_i=1, \tilde x_i),
    \quad\text{fitted on }\{i\in\{0,\ldots,n\}: y_i>0\}.
\end{equation}
\end{enumerate}

Here also, both learners are Random Forests, whose hyperparameters are calibrated in the same way as for the initial learners.
The final stacked prediction is:
\begin{equation}
\label{eq:final_prediction}
\hat y_i = \hat p_i\,\hat \mu_i.
\end{equation}

Predictive performance of the stacked hurdle model was evaluated using an inter-annual cross-validation procedure. For each target year, the model was trained on observations from all other years together with a variable proportion of observations from the target year, while the remaining observations from the target year were used for validation. Additional details on training and validation are provided in Appendix C.

\subsection{A spatio-temporal semi-parametric model for epidemic mapping}

In a second step, our objective is to reconstruct the epidemic dynamics at the regional scale from pointwise information on disease severity, typically the predicted severity values ($\hat{y_i}$), obtained from covariates using the stacked hurdle model described above. Here, we will also reconstruct the epidemic dynamics from observed severity values ($y_i$), as a benchmark for our approach.
The goal of this analysis is to reconstruct the spatio-temporal evolution of the epidemic in order to provide, over the course of the epidemic, a spatially continuous estimate of the average disease severity throughout the study region.
To this end, we model disease severity across space and time using a semi-parametric framework that combines a parametric temporal component with a non-parametric spatial smoother. Here, we suppose that the observation of covariates $x_i$ is {\it cheap} enough so that the predicted severity values $\hat{y_i}$ are available at multiple time points for, at least, a subset of locations.

\paragraph{Notation.}

We restrict the analysis to a given year. Let $\{u_j\}_{j=1}^m$ denote the set of unique spatial locations --or individuals-- with at least one severity value available in the data. Let ${\mathcal I}_j$ denote the set of indices $i\in\{1,\ldots,n\}$ such that $t_i$ corresponds to the year of interest and $s_i=u_j$. Moreover, let $N_j$ be the number of such indices, $N_j=\text{card}({\mathcal I}_j$). 

We then partition the set $\{u_j\}_{j=1}^m$ into frequently observed locations ($\mathcal{U}_{\mathrm{frequent}}$) and temporally sparsely observed locations ($\mathcal{U}_{\mathrm{sparse}}$):
\[
\mathcal{U}_{\mathrm{frequent}} = \{\, u_j : N_j \ge N^\star \,\},
\qquad
\mathcal{U}_{\mathrm{sparse}} = \{\, u_j : N_j < N^\star \,\},
\]
where $N^\star\in\mathbb N^*$ is a threshold value used to determine whether a disease progress curve is directly fitted to local observations (when $N_j \ge N^\star$) because there are enough data, or fitted in a hybrid manner by using both local observations and an initial estimate of the local disease dynamics (when $N_j < N^\star$).
The purpose of the semi-parametric model is to construct a spatio–temporal estimator $\hat y(s,t)$, $(s,t)\in \mathcal{S}\times \mathcal{T}$, that provides disease severity over a continuous space–time domain $\mathcal{S}\times \mathcal{T}$, where $\mathcal S$ covers the geographic region under study, and $\mathcal T$ covers the epidemic season of the year of interest.

\paragraph{Parametric disease progress curve.}

Disease progress over time at locations of interest is modelled using a logistic growth function \citep{werker_modelling_1998}, which captures a common pattern in infectious disease dynamics: an initial slow increase, followed by a phase of rapid growth, and an asymptotic plateau when observations extend over a sufficiently long period. This behaviour is consistent with our case study and, more generally, arises in SI and SIS compartmental models \citep{allen_discrete-time_1994}, as well as during the early to mid phases of SIR models \citep{brauer_compartmental_2008}, where S, I, and R denote susceptible, infectious, and removed compartments, respectively. Other shapes (including non-monotonous shapes) may be used depending on the application (see discussion in Section \ref{sec:discu}).

For each $u_j \in \mathcal{U}_{\mathrm{frequent}}\cup\mathcal{U}_{\mathrm{sparse}}$, the severity value at time $t \in \mathcal{T}$ of the disease progress curve (DPC), denoted $\tilde y(u_j, t)$, is given by:
\begin{equation}
\label{eq:dpc_function}
\tilde y(u_j,t)
= \frac{1}{1 + \exp\!\left(-r_j\,\big(t - t_j^{50}\big)\right)},
\end{equation}
where $r_j \in \mathbb R^{+*}$ denotes the disease progress rate at location $u_j$, and $t_j^{50}\in\mathbb R^{+}$ the inflection time at which severity reaches $50\%$.
The DPC parameters $\big(r_j,\, t_j^{50}\big)$ are location-specific to account for a wide heterogeneity in disease progress, especially in the starting times of local --or individual-- disease dynamics that can occur earlier or later in the epidemic season depending on the location.

\paragraph{DPC parameter estimation at frequently observed locations.}

The first step of the semi-parametric framework consists of estimating DPC parameters $(r_j,\, t_j^{50})$ at each frequently observed spatial location. This is done independently across locations $u_j$ in $\mathcal{U}_{\mathrm{frequent}}$, by minimizing the root mean squared error (RMSE) between available severity values and the values predicted by the logistic function:
\begin{align}
\label{eq:argmin_RMSE_dpc}
\big(\tilde r_j,\, \tilde t_j^{50}\big)
&=\underset{r>0,\,t^{50}\in\mathbb R^{+}}{\arg\min}\;
\mathrm{RMSE}_j(r,t^{50}),\\
\label{eq:RMSE_dpc}
\mathrm{RMSE}_j(r,t^{50})
&= \sqrt{ \frac{1}{N_j}
\sum_{i\in{\mathcal I}_j}
\left( \hat y_i - \tilde y(u_j, t_i)
\right) ^2}.
\end{align}

\paragraph{DPC parameter estimation at sparsely observed locations.}

For sparsely observed locations $u_j$ in $\mathcal{U}_{\mathrm{sparse}}$, DPC parameters are estimated using a weighted sum of two RMSE values:
\begin{equation}
\label{eq:weighted_refit}
(\tilde r_j,\, \tilde t_j^{50})
= \underset{r>0,\,t^{50}\in\mathbb R^{+}}{\arg\min}\;
 \left( \mathrm{RMSE}_j(r,t^{50}) + w_j \mathrm{RMSE}_j^{\text{smooth}}(r,t^{50}) \right).
\end{equation}
In \eqref{eq:weighted_refit}, $w_j$ is a positive weight to be selected in conjunction with $N_j$ (see Appendix C), $\mathrm{RMSE}_j(r,t^{50})$ is the data-based RMSE given by Equation \eqref{eq:RMSE_dpc}, and $\mathrm{RMSE}_j^{\text{smooth}}(r,t^{50})$ is the smoother-based RMSE between (i) a local smoother of disease progress computed from DPCs fitted at frequently observed locations and (ii) the logistic function given by Equation \eqref{eq:dpc_function}. 
More precisely,
\begin{equation}
\mathrm{RMSE}_j^{\text{smooth}}(r,t^{50}) = 
\sqrt{  \frac{1}{L}
\sum_{\ell=1}^{L}
\bigl[ \hat y^{(0)}(u_j,\tau_\ell) - \tilde y (u_j,\tau_\ell) \bigr]^2 },
\end{equation}
where $\tau_1<\cdots<\tau_L$ form a regular grid over the time period $\mathcal{T}$, and $\hat y^{(0)}(\cdot,\tau_\ell)$ is a Nadaraya-Watson spatial smoother \citep{lawson_statistical_2013} of estimated DPC values obtained at frequently observed locations at time $\tau_\ell$. Thus, for all $u_j\in{\mathcal U}_\text{sparse}$,
\begin{equation}
\label{eq:kernel_smoothing_sites}
\hat{y}^{(0)}(u_j,\tau_\ell)
= \frac{
\displaystyle\sum_{u \in \mathcal{U}_{\mathrm{frequent}}}
K\!\left(\frac{d(u_j, u)}{h^{(0)}}\right)\, \hat{\tilde y}(u,\tau_\ell)}
{\displaystyle\sum_{u \in \mathcal{U}_{\mathrm{frequent}}}
K\!\left(\frac{d(u_j, u)}{h^{(0)}}\right)} ,
\end{equation}
where $d(u_j, u)$ denotes the Euclidean distance between $u_j$ and $u$, $h^{(0)}$ is the spatial bandwidth (selected by cross-validation), $K$ is the Gaussian kernel, 
$K(u) = (2\pi)^{-1/2}\exp(-u^2/2)$, and $\hat{\tilde y}$ is obtained by plugging the estimates $(\tilde r_j,\, \tilde t_j^{50})$ given by Equation \eqref{eq:argmin_RMSE_dpc} in $\tilde y$ expressed in Equation \eqref{eq:dpc_function}.

Hence, to fit the DPC at sparsely observed locations, we complement the limited available local data with information on the DPC shape derived from neighbouring, more frequently observed locations. Indeed, the pairs $(\tau_\ell,\hat{y}^{(0)}(u_j,\tau_\ell))$, $\ell=1,\ldots,L$, play the role of pseudo-observations at location $u_j\in{\mathcal U}_\text{sparse}$, and the weight $w$ can be tuned to determine how much these pseudo-observations influence the estimation of DPC parameters with respect to actual data $\{(t_i,\hat y_i):i\in{\mathcal I}_j\}$.

\paragraph{Semi-parametric estimation of disease spread.}

The previous paragraphs lead to estimates of the DPC at both frequently and sparsely observed locations. For sparsely observed locations, we actually introduced an initial estimator of disease severity across space and time, namely $\hat y^{(0)}$ given by Equation \eqref{eq:kernel_smoothing_sites}, leveraging DPC estimates at frequently observed locations. 
Here, we propose a similar estimator of disease spread, which exploits DPC estimates at both frequently and sparsely observed locations to reconstruct the spatial distribution of disease severity over the entire spatial domain for the different dates considered.

Thus, for all location $s\in\mathcal S$ and time $t\in\mathcal T$, disease severity is estimated by:
\begin{equation}
\label{eq:kernel_smoothing}
\hat{y}(s,t)
= \frac{
\displaystyle\sum_{u \in \mathcal{U}_{\mathrm{frequent}}\cup \mathcal{U}_{\mathrm{sparse}}}
K\!\left(\frac{d(s, u)}{h}\right)\, \hat{\tilde y}(u,t)}
{\displaystyle\sum_{u \in \mathcal{U}_{\mathrm{frequent}}\cup \mathcal{U}_{\mathrm{sparse}}}
K\!\left(\frac{d(s, u)}{h}\right)} ,
\end{equation}
where $d$ and $K$ were already defined for Equation \eqref{eq:kernel_smoothing_sites}, $h$ is the spatial bandwidth (selected by cross-validation, possibly different from $h^{(0)}$), and $\hat{\tilde y}$ is obtained by plugging the estimates $(\tilde r_j,\, \tilde t_j^{50})$ given by Equation \eqref{eq:argmin_RMSE_dpc} in $\tilde y$ expressed in Equation \eqref{eq:dpc_function} for $u\in\mathcal{U}_{\mathrm{frequent}}$, and the estimates $(\tilde r_j,\, \tilde t_j^{50})$ given by Equation \eqref{eq:weighted_refit} in $\tilde y$ for $u\in\mathcal{U}_{\mathrm{sparse}}$.

Hence, Equation \eqref{eq:kernel_smoothing} yields smooth epidemic surfaces that represent the average disease severity for any location $s$ and time $t$, combining reconstructed temporal dynamics, and spatial information from all available locations.

\section{Results}\label{sec:results}

\subsection{Stacked hurdle model and prediction of yellows severity at the field level}

\begin{figure}
\centering
\includegraphics[width=17cm]{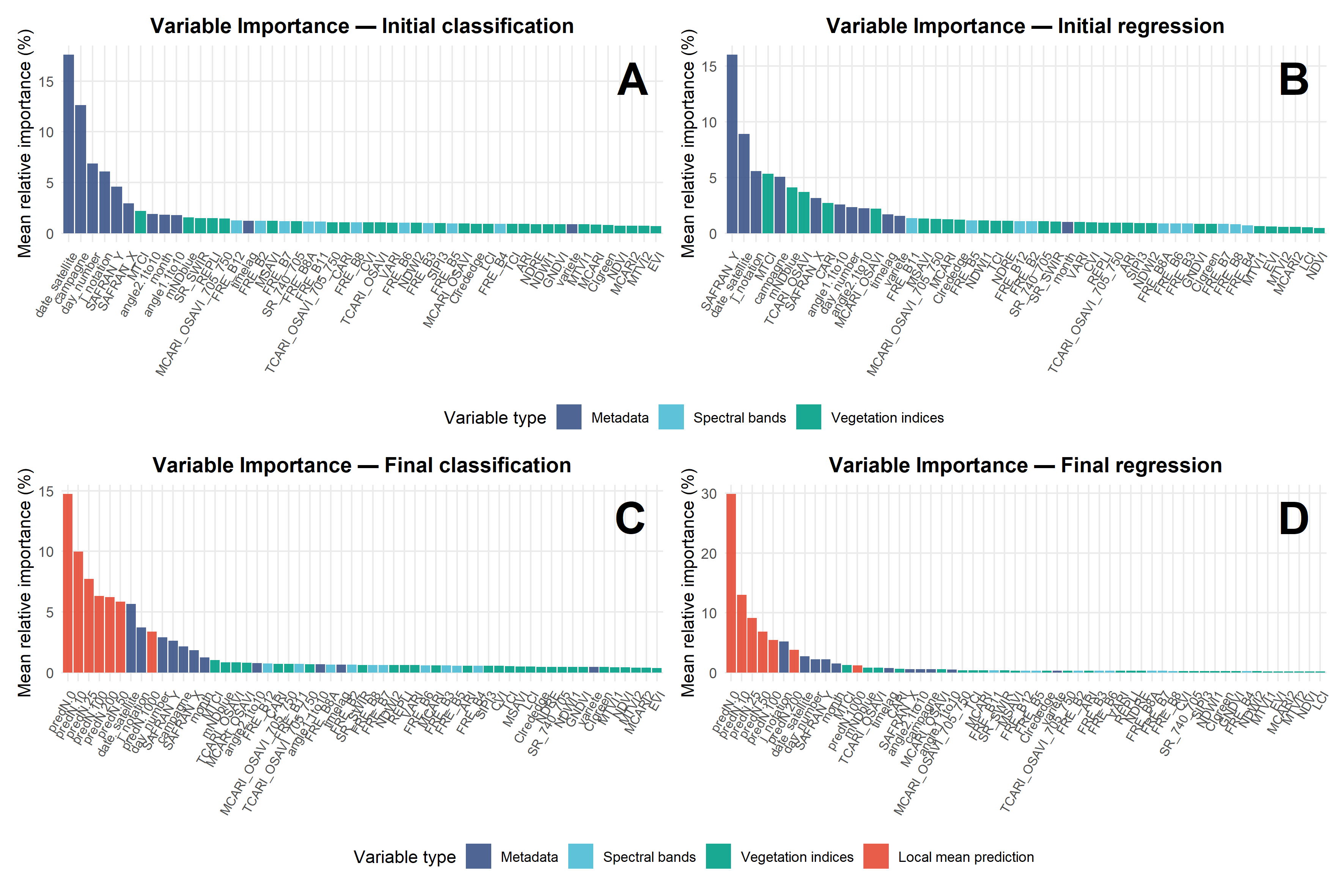}
\caption{\label{fig:var_importance} Each panel represents the mean relative importance (\%) of each covariate for initial learner RF classification, (B) initial learner RF regression, (C) final learner RF classification, and (D) final learner RF regression. Results are expressed as the average importance across 40 runs (4 years, 10 repetitions, with a varying proportion of target-year data included in training).}
\end{figure}

Figure~\ref{fig:var_importance} summarizes the relative importance of covariates across the four Random Forest models used in the field-scale prediction framework. As expected, since beet yellows dynamics are structured in both space and time, metadata related to the spatial position ($SAFRAN\_X$, $SAFRAN\_Y$) and temporal context of the satellite acquisitions ($date\_satellite$, $j\_notation$, $day\_number$, $month$) are dominant predictors in both the initial classification and regression Random Forests. 
Vegetation indices, computed from raw spectral bands, also contribute substantially, particularly those related to chlorophyll absorption ($MTCI$, $mNDblue$, $TCARI OSAVI$, $CARI$, $MCARI$ $OSAVI$). Raw spectral bands contribute more moderately but consistently across models, with slightly higher importance for blue ($FRE\_B2$), red-edge ($FRE\_B5$, $FRE\_B7$) and shortwave-infrared bands ($FRE\_B11$, $FRE\_B12$).
For the final learners, similar patterns are observed regarding the dominance of spatio-temporal metadata and spectral information. In addition, the meta-features, computed as mean predicted values from the initial learner over different spatial neighbourhoods, play a central role in both the classification and regression frameworks.  The importance of these predictors decreases with increasing neighbourhood size. In particular, the local prediction ($predN.0$) and the averages computed within 10 ($predN.10$) and 25 km ($predN.25$) neighbourhoods emerge as the most influential among these spatially aggregated covariates.

\begin{figure}
\centering
\includegraphics[width=15cm]{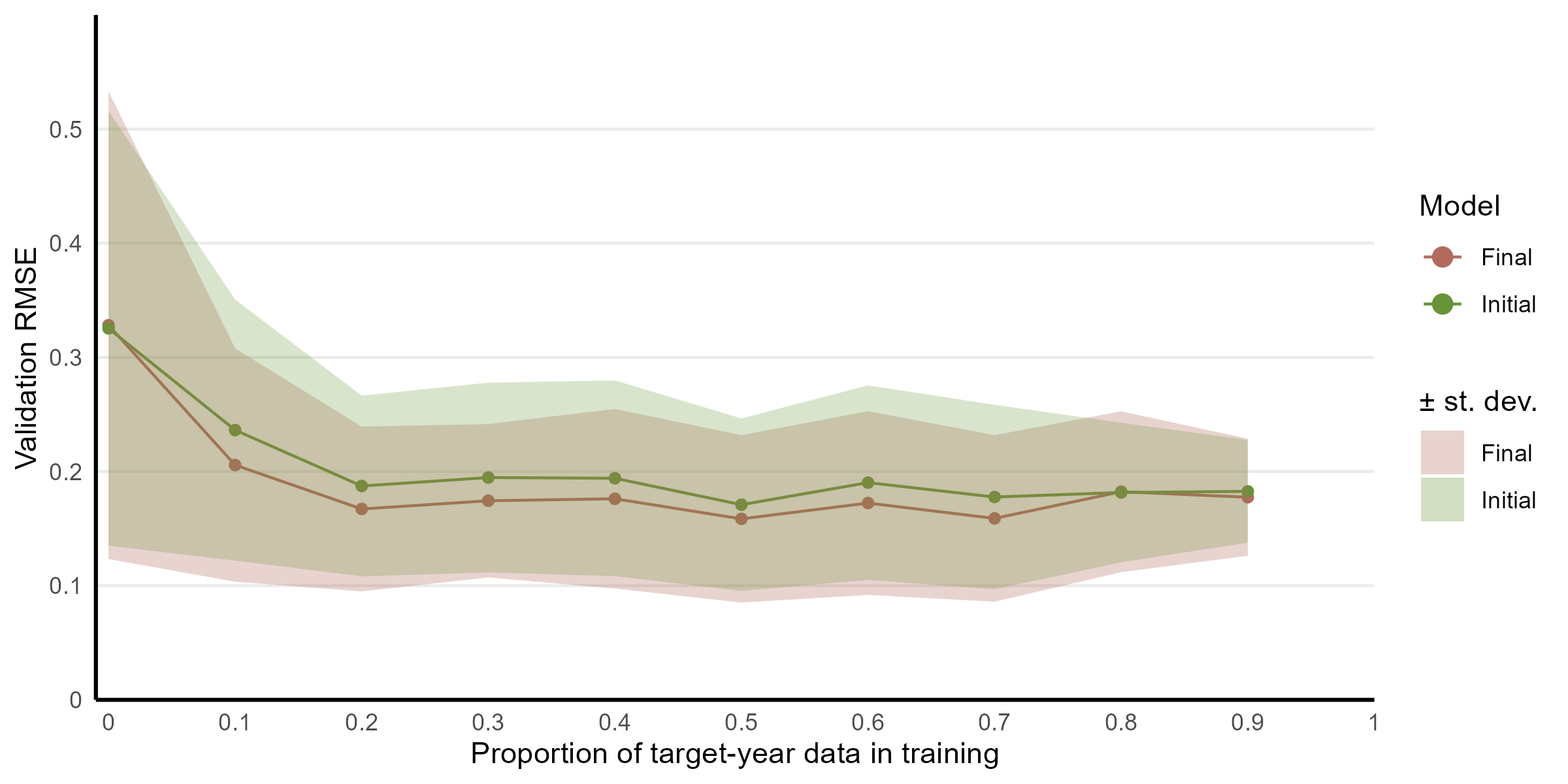}
\caption{\label{fig:rmse_prop} RMSE validation ---for both initial and final learner--- versus the proportion of data from the predicted year included in the training dataset. For each proportion, RMSE values are weighted by the number of predictions in each year and averaged across 2019, 2020, 2021 and 2023. The shaded ribbons show the weighted standard deviation, based on the same weighting scheme.}
\end{figure}

The performance of the models used to predict beet yellows severity from the available data is displayed in Figure~\ref{fig:rmse_prop}, and detailed for each year in Appendix D. Figure~\ref{fig:rmse_prop} highlights how the proportion of target-year data included in the training dataset impacts prediction quality. 
RMSE sharply decreases when this proportion increases from 0.0 to 0.2, and remains approximately constant when the proportion of target-year data further increases. This effect is particularly pronounced for 2020, which presented a unique epidemiological scenario not represented in the training data drawn from other years (Appendix D).

Figure~\ref{fig:rmse_prop} also shows that the RMSE of the final model are approximately 10\% lower than those of the initial model as soon as target-year data are integrated ($proportion > 0$). This indicates the added value of the stacked modelling strategy.

\begin{figure}
\centering
\includegraphics[width=16cm]{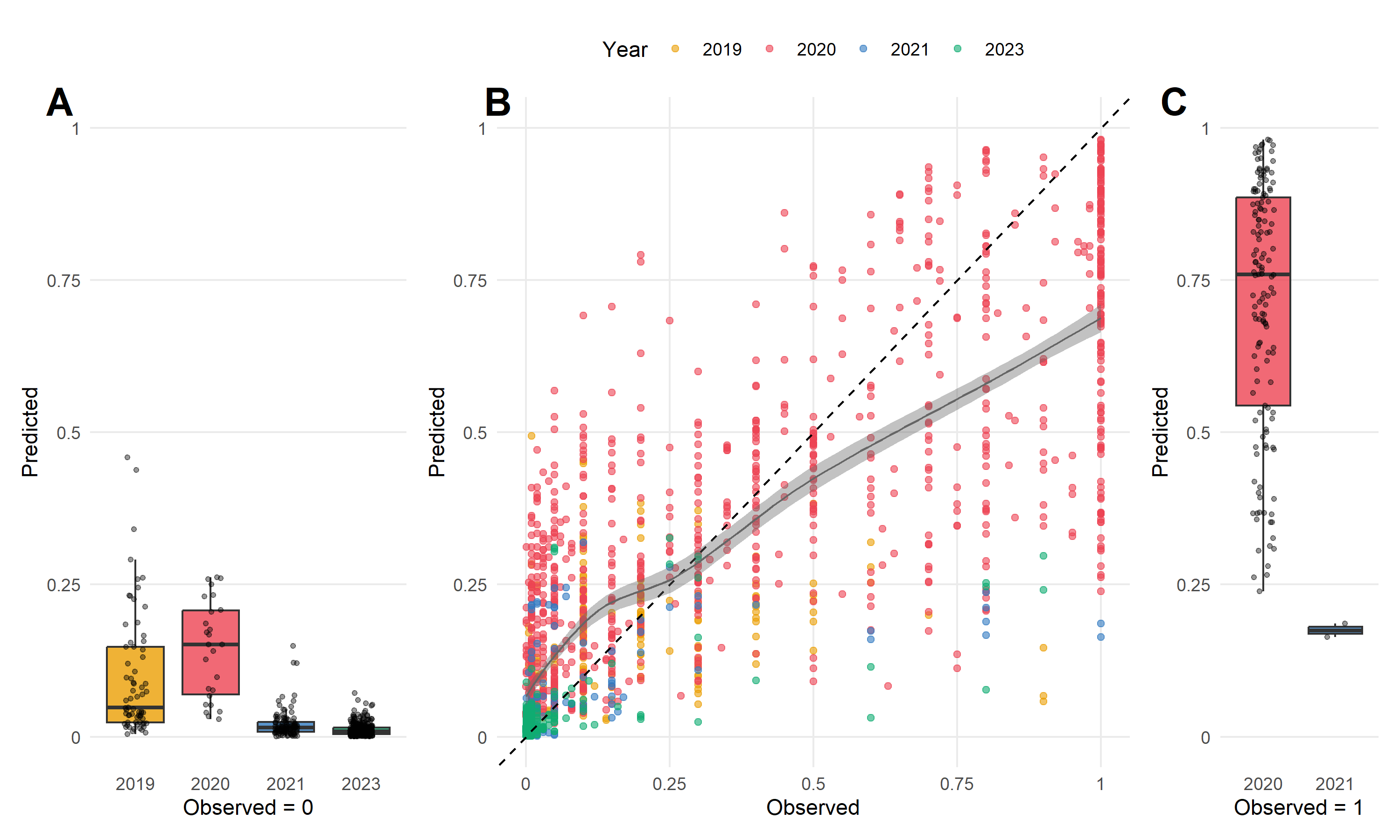}
\caption{\label{fig:obs_vs_pred} Comparison of observed values with predicted values on validation data when 20\% of target-year data are included in training. A: Predicted values where actual observed severity is zero, classified by year. B: Predicted values where actual severity is ranged between 0 and 1, with 0 excluded (grey curve and associated envelope: LOESS smoother and 95\% pointwise confidence envelope). C: Predicted values for observations where the observed severity is equal to one.}
\end{figure}

Figure ~\ref{fig:obs_vs_pred} compares the values of yellows severity predicted in a cross-validation setting with the observed values. Results show a reasonable agreement between predicted and observed severity, despite a substantial variability in prediction that reflects the limited ability of the model to accurately reproduce severity at the individual field level in any situation. The model generally succeeds in identifying the presence of yellows at individual fields, but tends to produce some false negatives (mostly associated with low predicted severities in the subsequent regression model). Overall, the model is able to distinguish between fields with little or no infection and those that are strongly affected by beet yellows. The hurdle structure of the model contributes to this performance: during the classification stage, the probability of assigning a field to the non-zero severity class is higher than 0.75 for almost all fields exhibiting moderate or high observed severity (see Appendix E).

The largest number of observations and infected fields is recorded in 2020, and this year largely drives the overall trend observed in Figure~\ref{fig:obs_vs_pred}. Given the high disease level that year, the model tends to overestimate the presence of yellows in fields with only mild symptoms. A large dataset is also available in 2023 (Figure \ref{fig:illus_data}), but in contrast to 2020 the severity of yellows was low during that year. Results show that model prediction accuracy is high in this context. Fewer observations were available in 2019 and 2021, and in that setting the model generally succeeds in identifying the presence/absence of yellows, but the regression model is not particularly accurate for predicting severity in infected fields as already observed.

\subsection{Epidemiological dynamic modelling}

\begin{figure}
\centering
\includegraphics[width=16cm]{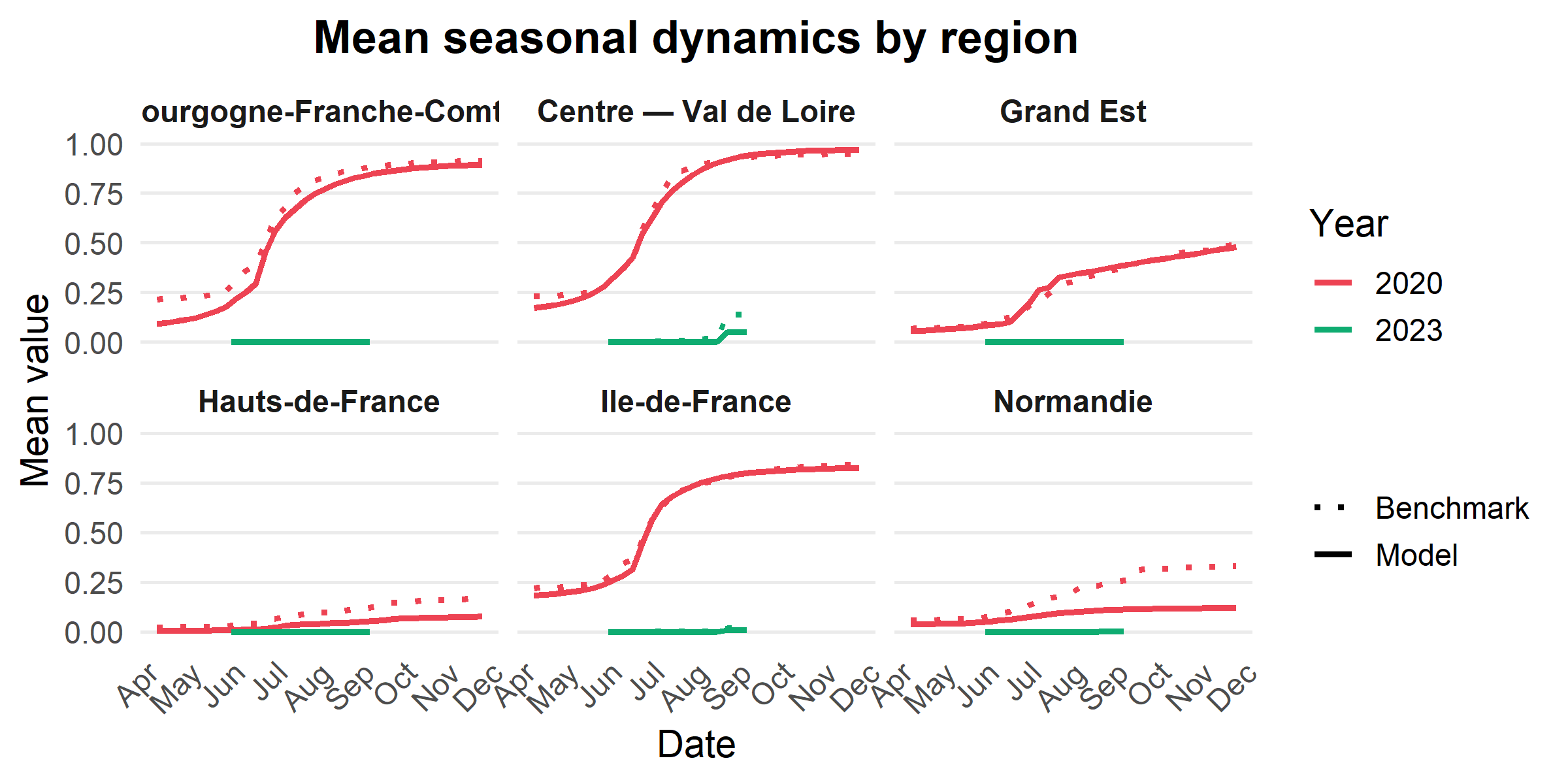}
\caption{\label{fig:season_dynamics} Temporal evolution of sugar beet yellows severity predicted by the spatio-temporal semi-parametric model for the 2020 and 2023 seasons averaged for all 6 regions presented in  Figure \ref{fig:illus_data}. Values correspond to regional mean severity estimates averaged over all pixels within each region. Solid lines correspond to the satellite-based, smoothed spatio-temporal severity estimates, while dotted lines correspond to the benchmark derived from field observations. Disease-progress curves are fitted over the full observation period of each season.}
\end{figure}

In 2020, a year marked by widespread beet yellows across much of the territory, the reference curves produced as the benchmark based on direct observations show the rapid increase of yellowing severity from June onwards, which began to plateau around August in in Bourgogne–Franche-Comté, Centre–Val de Loire, and Île-de-France, which were the most affected regions; see Figure \ref{fig:season_dynamics}. For these regions, the model, trained with 20\% of target-year data reliably reproduces the temporal dynamics of beet yellows. In the three other regions, which were less affected by the disease, the model captures the overall trend but slightly underestimates it in Normandy and Grand Est. These regions being located at the outskirt of the study area, this underestimation may come from edge effects of the spatial smoothing procedure. Overall, the epidemiological model calibrated using satellite-based predictions provides a reliable depiction of disease dynamics that year (Appendix F).

In 2023, four of the six regions exhibited no significant beet-yellows severity level. In this scenario, both the spatio-temporal model and the benchmark appropriately yield near-zero severity estimates. Only Centre–Val de Loire and Île-de-France show low levels of disease. Here again, the model successfully reproduced the general epidemic pattern, with little to no disease, with a slight underestimation in Centre-Val de Loire. These results are illustrated by the maps provided in Appendix F.

The RMSE curves comparing the model outputs with the benchmark are consistent with the findings above (Appendix G). Beyond the uncertainty associated with the stacked hurdle model, the semi-parametric model is accurate for both high-severity and low-severity areas. The bootstrap-derived standard errors were close to zero in these areas (Appendix H). Intermediate-severity areas were also generally well captured, though with slightly higher associated uncertainty. Overall, these findings show that the approach remains stable under markedly different signal-to-noise conditions.

\section{Discussion}\label{sec:discu}

We proposed a general and modular statistical framework for inferring disease severity and reconstructing its spatio-temporal spread from indirect indicators, with potential applications across human, animal, and plant epidemiology. The stacked hurdle model combined with the semi-parametric spatio-temporal model provides a coherent, modular, interpretable, and general-purpose framework for large-scale epidemic monitoring. The application to the plant epidemiology case study presented in this article illustrates both the predictive potential of the method and its ability to infer disease severity values from a reference dataset reduced by a factor of five. Below, we discuss epidemiological insights and modelling choices, envisioning possible extensions of our work.

\subsection{Insights from the sugar beet yellows motivating problem}

Figures \ref{fig:rmse_prop} \& \ref{fig:obs_vs_pred} highlight the relevance of a stacked hurdle model for predicting beet yellows severity. The model is able to separate fields with no or weak symptoms from those more strongly affected by the disease and succeeds in dealing with zero-inflation. From an agronomic perspective, the relative importances given to the different covariates are consistent with existing knowledge on beet yellows occurrence and spread (see Appendix I), granting credibility to the stacked hurdle model outputs. The uncertainty of field-scale predictions remains significant due to a combination of factors: the limited spatial resolution of available images (the spatial resolution of the Sentinel‑2 bands used in this study is 10m or 20m); the possible existence of confounding factors such as water stress \citep{yetik_chlorophyll_2023}, nitrogen status \citep{elsayed_estimating_2023} or fungal diseases \citep{mahlein_spectral_2010}; the effect of heterogeneous management practices as well as the spatially-varying plant phenology; and imperfect field-level reference observations, which were assessed visually.

The sensitivity of prediction accuracy to the inclusion of target-year observations (Figure \ref{fig:rmse_prop}) underscores the pronounced inter-annual variability of beet yellows epidemics, already reported in the literature \citep{hossain_new_2021, mahillon_virus_2022}. As a result, including even small proportions of target-year data in the training dataset produces substantial gains in accuracy, indicating that the statistical relationships learned by the model are not fully transferable from one year to the other. This highlights the complementarity between the broad coverage of remote sensing data and the more specific information carried by field observations, which remain essential for anchoring the target-year disease expression into the model. This result is also encouraging from an operational perspective because it implies that satisfactory predictions of field-scale beet yellows severity can be achieved with an observation effort reduced by 80\%. Moreover, as most fields are currently unobserved, the model could offer a new way to monitor severity at a larger scale, typically a production basin. However this requires to know where beet fields are located, which is often known only after the growing season through agricultural declarations. This confines direct applications to the retrospective assessment of yellows-induced damages and yield losses, for instance for insurance purposes. Large-scale real-time monitoring of sugar beet severity is expected to be more challenging because it would require to train the model exclusively on observations collected before the forecast date.

The semi-parametric spatio-temporal model yields coherent epidemic dynamics across the two intensively-monitored years 2020 and 2023 despite (i) contrasted epidemiological contexts, and (ii) variable satellite-based severity estimates. In 2020, a high-severity year, the model based on predictions including remote sensing data closely reproduced the temporal dynamics inferred from the benchmark. In 2023, a year characterised by a low disease prevalence, errors were close to zero early in the season and remained low afterwards. This is because the model did not generate false positives in this type of scenario, at the expense of an underestimation of yellows severity in Eure-et-Loir compared to the benchmark. Overall, these results demonstrate the potential of coupling the model with remote sensing data to reconstruct a posteriori the epidemic dynamic in order to better understand the spread of the disease in space and time. Future steps in this direction include exploiting all available acquisition dates rather than only those coinciding with field observations, and integrating additional explanatory data in the analysis for instance meteorological conditions, bioclimatic variables, land-use information, or management practices \citep{chauvin_factor_2025}.

\subsection{Methodological considerations and perspectives} 

The originality of the proposed approach relies on two key features:

 \begin{itemize}
\item \textit{A semi-parametric design.}  
The proposed approach allows a balance between model complexity and underlying simplifying assumptions that can be tailored depending on the amount of available training data and their information content. The modelling approach should be robust in data-limited conditions while remaining sufficiently flexible to capture complex epidemiological patterns.

\item \textit{Strong modularity.}  
Base learners, here based on random forest, can be easily replaced to better match the properties of the available indirect indicators (remote sensing, diagnostics, syndromic signals, etc.) and exploit the sample size. The number of stacked layers is also flexible, enabling the model to capture multi-scale spatio-temporal patterns or to accommodate specific data characteristics, as illustrated here by the explicit treatment of zero inflation. Importantly, local severity estimation is decoupled from large-scale epidemic reconstruction, whose temporal and spatial components can be adapted to the epidemiological context in which the framework is deployed.

\end{itemize}

The first stage of the framework proposed in this article relies on hurdle model factorisations, that yield good results when applied to zero-inflated data as reported previously in the literature \citep{feng_comparison_2021, rozanec_dealing_2025}. We also evaluated alternative modelling strategies that directly account for zero inflation within a single model. In particular, a GAMLESS model combined with a beta-inflated distribution produced competitive results, although it remained slightly less performant than the hurdle random forest approach adopted here. The use of a final learner based on neighbourhood meta-features derived from the initial learner contributes with limited but systematic performance gains compared to an architecture that ignores spatial structure. This result is consistent with the intuition that local spatial context conveys relevant epidemiological information. The stacked hurdle model is therefore able to leverage spatial structure in a conceptually simple and modular way, extending ideas previously explored in spatio-temporal hurdle modelling approaches for other types of zero-inflated spatial processes \citep{sales_spatiotemporal_2017}. It should be noted, that since the meta-features are based on the predictions of the initial hurdle model, there is a risk of introducing bias into the meta-features and causing overfitting in the final model. This could also make the model less generalizable for temporal transfer, which could partly explain the need to include a proportion of data from the target year in the training set. Interestingly, the structure of the stacked hurdle model can be represented as a heterogeneous layered ensemble, sharing structural similarities with neural networks (see Appendix J). Neural-network-based architectures were also explored and yielded comparable predictive performance, but at the cost of reduced interpretability.

The semi-parametric spatio-temporal architecture is particularly well suited to reconstruct epidemic dynamics at the scale of a geographical region, because it combines an epidemiologically meaningful temporal structure with a flexible description of the spatial heterogeneity. In this framework, the logistic DPC provides an interpretable temporal base capturing onset, epidemic growth, and saturation in a simple and consistent way. The spatial smoothing step then absorbs local inconsistencies without imposing a rigid spatial correlation structure. Application to the case study yields good results and show that the approach remains stable under different signal-to-noise conditions. 

Depending on the application, alternative models could accommodate more complex temporal patterns than a logistic DPC, nevertheless requiring a denser temporal sampling. For instance, curves resulting from more sophisticated compartmental, epidemic models \citep{bjornstad_seirs_2020}; spline-based growth curves gives local flexibility \citep{syarif_corn_2018}; generalized logistic or Richards curves capture varying inflection points \citep{alves_analysis_2021, xu_modelling_2006}; and fully non-parametric approaches (e.g., Gaussian processes,  neural network), which can model arbitrary shapes \citep{ince_non-parametric_2006}. Regarding the spatial dimension, the choice of a spatial smoothing approach, rather than methods that explicitly model the spatial structure of disease severity, is motivated by the objective pursued in this study. Mapping epidemic dynamics at the scale of the production region requires approaches that are robust and weakly sensitive to uncertainty in the input data, such as data originated from predictive models. In this context, weakly parametric and stable smoothing methods are preferred. A Gaussian kernel was selected as a simple smoothing function, providing a continuous and isotropic decay of spatial influence without requiring explicit spatial covariance structure. More explicit spatial modelling strategies could also be considered depending on the objectives pursued. Mechanistic spatio-temporal epidemic models could, for instance, more explicitly represent transmission dynamics and spatial interactions \citep{mahmood_modeling_2022}. Classical geostatistical alternatives such as ordinary or universal kriging, Gaussian random fields, or spatial GAMs could provide richer spatial, at the cost of stronger assumptions on spatial covariance and increased modelling complexity \citep{auchincloss_review_2012, bonsoms_comparison_2024}. In preliminary experiments, such approaches were also evaluated but proved more sensitive to prediction errors from the stacked hurdle model.

\subsection{Conclusion}

Overall, results obtained on the case-study highlight the strong potential of combining indirect indicators, such as remote sensing data, with the proposed modelling framework for epidemiological monitoring at large spatial scales. As data volumes increase and these indirect indicators become increasingly available, this type of model could become central to (human, animal and plant) epidemiology, where direct observations of disease are often limited and costly. When larger datasets become available, future work could explore approaches integrating the temporal structure more explicitly in the stacked hurdle model, for example by directly incorporating disease progress curve dynamics or temporal trajectory descriptors in the learning process. Replacing random forest with more complex machine learning such as deep neural networks \citep{wilson_beyond_2022} could also be advantageous in contexts involving very large training datasets and complex, high-dimensional predictors.

\vspace{1em}

\noindent\textbf{Funding.}
This work was supported by the BEET-SAT project (CASDAR -- FranceAgriMer Grant 21611650), the SEPIM project (FranceAgriMer Grant 3890396), and the BEYOND project (Grant ANR-20-PCPA-0002).

\vspace{1em}

\noindent\textbf{Acknowledgements.}
The authors thank David Sheeren and Corentin Barbu for their helpful comments on the manuscript. They also thank the stakeholders involved in the PNRI and the PNRI-C programmes (“France Relance” initiative; Agriculture, Food and Forest Transition) for fruitful discussions and access to data, in particular Fabienne Maupas from ITB. This work is the sole responsibility of the authors; the French Ministry of Agriculture, although involved in its funding, cannot be held responsible for its content.

\vspace{1em}

\noindent\textbf{Data and code availability.}
The code implementing the modelling framework presented in this article, together with the data associated with the sugar beet yellows case study, are available at: \url{https://doi.org/10.5281/zenodo.20311758} . The main code is also available on GitHub at: \url{https://github.com/Baptiste-Oger/Disease_Predict_Spread}.

\clearpage

\appendix

\renewcommand{\thefigure}{S\arabic{figure}}
\renewcommand{\thetable}{S\arabic{table}}
\renewcommand{\theequation}{S\arabic{equation}}

\setcounter{figure}{0}
\setcounter{table}{0}
\setcounter{equation}{0}

\section*{Supplementary Material}
\addcontentsline{toc}{section}{Supplementary Material}

\section{Extended background on sugar beet yellows monitoring}

\subsection{Sugar beet yellows}

Sugar beet is the second most important crop for sugar production in the world, after sugarcane, and the most important in temperate regions. During the 2023/2024 season, an estimated 15.3 million tons of sugar beet were produced in Europe, where it constitutes the main source of sugar \citep{cefs_european_2025}. Beet yellows disease represents the main threat to sugar beet production. These diseases are associated with three main viruses, the Beet Mild Yellowing Virus (BMYV), the Beet Chlorosis Virus (BChV), and the Beet Yellows Virus (BYV), which are all transmitted by aphids. The disease induced by these three viruses causes yellowing of the affected leaves, thereby limiting photosynthetic activity. This, in turn, results in stunted growth and reduced sugar content. For farmers, yield losses can reach 20\% to 40\%, depending on the viral strain, with the most severe impacts being associated with BYV \citep{clover_effects_1999, hossain_new_2021}. Although chemical seed treatments with neonicotinoids substantially reduced the impact of beet yellows for several decades, numerous studies have since highlighted their harmful effects on human health \citep{zhang_potential_2018}, pollinator populations \citep{blacquiere_neonicotinoids_2012}, as well as their persistence in plants and soils\citep{jones_neonicotinoid_2014, viric_gasparic_neonicotinoid_2020}. These concerns led to restrictions across Europe from 2018 onwards, coinciding with a resurgence of sugar beet yellows and significant yield losses in several regions \citep{verheggen_producing_2022}. In this context, improving the capacity to monitor beet yellows at large spatial scales and throughout the entire growing season has become a priority for farmers, advisers, and plant-health services. Early detection of disease severity enable timely prophylactic interventions, to prevent disease propagation and worsening, and provide objective information on potential yield-loss.

\subsection{Linking field yellows monitoring and satellite imagery}

Traditional methods for monitoring and assessing the severity of beet yellows largely rely on field observations carried out by farmers or other stakeholders, as well as on sample collection and laboratory testing when viral strain identification is required. These methods remain limited and too costly to enable precise tracking of the development and spread of the disease across space and time at the regional to country scale. At the same time, the recent expansion of remote sensing applications in agriculture offers new perspectives for crop monitoring \citep{wu_challenges_2023} and epidemiological modelling  \citep{mikaberidze_opportunities_2025}. In particular, multispectral satellite imagery, such as the products provided by the Copernicus and PlanetScope programs, make it possible to monitor beet yellows development in space and time over large areas. Recent studies have demonstrated the potential of these data for monitoring other sugar beet pests \citep{hillnhutter_remote_2011} as well as for evaluating  other crop parameter such as Leaf Area Index
\citep{richter_experimental_2009} or yield \citep{ludewig-spickermann_optimisation_2025, bouasria_use_2021}. In addition, remote sensing has successfully been used to detect virus-induced yellowing in other crops \citep{d_k_das_spectral_2013, guo_recognition_2022, zibrat_detection_2024}. However, no remote-sensing-based framework exists yet for detecting or monitoring beet yellows specifically, despite the suitability of current satellite products for this purpose. 

More generally, there is a growing need for statistical frameworks capable of exploiting multispectral time-series data to track foliar disease dynamics at both field and landscape scales. Such frameworks, when applied to beet yellows, raise several methodological challenges that echo those encountered more broadly in epidemiological surveillance, in particular: 1) The distribution of beet yellows severity data is strongly zero-inflated. Indeed, most fields exhibit no visible symptoms, especially during low-incidence years or when monitoring is conducted preventively early in the season. 2) Satellite-derived indicators provide only indirect and imperfect information on disease severity, requiring statistical models that are robust to noisy and incomplete inputs. For instance reflectance measurements are affected by meteorological conditions, missing acquisitions (i.e. cloudy images), and artefacts, and their relationship with yellows severity is often confounded by agronomic heterogeneity across fields. In addition, management practices, soil properties, and microclimatic conditions are often unknown factors that can influence crop light reflectance and disease expression \citep{zarco-tejada_remote_2024}. 3) Beet yellows epidemics are structured in both space and time. By feeding on infected plants, aphids acquire the yellow related viruses and subsequently transmit them to other plants within the same field and to nearby fields, with transmission being limited by aphid flight distance. This process results in spatially and temporally progressive patterns of disease severity, both within and between fields. 4) Field observations are costly and epidemiological monitoring frameworks should be able to leverage a small and potentially variable proportion of reference observations to assess severity across space and time for the entire target area covered by satellite imagery.

\section{List of covariates}

Table \ref{tab:covariates} lists all covariates used in the case study.

\begin{table}[htbp]
\small
\centering
\caption{List of covariates used in the Random Forest models, their type, and a short description or formula (for vegetation indices). In the formulas, $B_2, B_3, \ldots$ designate the spectral bands \textit{FRE\_B2}, \textit{FRE\_B2}, etc.}
\label{tab:covariates}
\begin{tabular}{llp{9.5cm}}
\hline
\textbf{Variable name} & \textbf{Type} & \textbf{Description / Formula} \\
\hline
\textit{campagne} & Metadata & Year of observation \\
\textit{variete} & Metadata & Sugar beet variety \\
\textit{SAFRAN\_X} & Metadata & Field longitude (from SAFRAN grid) \\
\textit{SAFRAN\_Y} & Metadata & Field latitude (from SAFRAN grid) \\
\textit{date\_satellite} & Metadata & Sentinel-2 acquisition date \\
\textit{month} & Metadata & Month of observation \\
\textit{day\_number} & Metadata & Date of observation expressed as the day number of the current year \\
\textit{j\_notation} & Metadata & Date of observation expressed in Julian days \\
\textit{timelag} & Metadata & Time difference (days) between satellite image and field observation \\
\textit{angle1.1to10} & Metadata & Satellite metadata zenith/azimuth angle for raw band i \\
\textit{angle2.1to10} & Metadata & Satellite metadata zenith/azimuth angle for raw band i \\[4pt]

\textit{FRE\_B2} & Spectral band & Sentinel-2 band 2 (490 nm, Blue) reflectance \\
\textit{FRE\_B3} & Spectral band & Sentinel-2 band 3 (560 nm, Green) reflectance \\
\textit{FRE\_B4} & Spectral band & Sentinel-2 band 4 (665 nm, Red) reflectance \\
\textit{FRE\_B5} & Spectral band & Sentinel-2 band 5 (705 nm, red-edge) reflectance \\
\textit{FRE\_B6} & Spectral band & Sentinel-2 band 6 (740 nm, red-edge) reflectance \\
\textit{FRE\_B7} & Spectral band & Sentinel-2 band 7 (783 nm, red-edge) reflectance \\
\textit{FRE\_B8} & Spectral band & Sentinel-2 band 8 (842 nm, NIR) reflectance \\
\textit{FRE\_B8A} & Spectral band & Sentinel-2 band 8A (865 nm, NIR narrow) reflectance \\
\textit{FRE\_B11} & Spectral band & Sentinel-2 band 11 (1610 nm, SWIR) reflectance \\
\textit{FRE\_B12} & Spectral band & Sentinel-2 band 12 (2190 nm, SWIR) reflectance \\[4pt]

\textit{NDVI} & Vegetation index & $(B8 - B4) / (B8 + B4)$ \\
\textit{VARI} & Vegetation index & $(B3 - B4) / (B3 + B4 - B2)$ \\  
\textit{EVI} & Vegetation index & $2.5 \times (B8A - B5) / (B8A + 6 \times B5 - 7.5 \times B2 + 1)$ \\ 
\textit{NDRE} & Vegetation index & $(B8A - B5) / (B8A + B5)$ \\
\textit{mNDblue} & Vegetation index & $(B2 - B5) / (B2 + B8)$ \\ 
\textit{MTCI} & Vegetation index & $(B7 - B5) / (B5 - B4)$ \\ 
\textit{TCARI\_OSAVI} & Vegetation index & $3[(B5 - B4) - 0.2(B5 - B3)(B5/B4)] / [1 + 0.16(B8 - B4)/(B8 + B4 + 0.16)]$ \\ 
\textit{TCARI\_OSAVI\_705\_750} & Vegetation index & $[3 * (B7 - B5) - 0.2 * (B7 - B3) * B7 / B5] / [(1 + 0.16) * (B7 - B5) / (B7 + B5 + 0.16)]$  \\ 
\textit{MCARI\_OSAVI} & Vegetation index & $[(B5 - B4) - 0.2(B5 - B3)(B5/B4)] / [(B8 - B4)/(B8 + B4 + 0.16)]$ \\ 
\textit{MCARI\_OSAVI\_705\_750} & Vegetation index & $[(B7 - B5) - 0.2 * (B7 - B3) * B7 / B5] / [(1 + 0.16) * (B7 - B5) / (B7 + B5 + 0.16)]$ \\ 
\textit{CIgreen} & Vegetation index & $(B7 / B3) - 1$ \\ 
\textit{CIrededge} & Vegetation index & $(B7 / B5) - 1$ \\ 
\textit{NDWI1} & Vegetation index & $(B8 - B11) / (B8 + B11)$ \\
\textit{NDWI2} & Vegetation index & $(B8 - B12) / (B8 + B12)$ \\
\textit{SR\_SWIR} & Vegetation index & $B11 / B12$ \\  
\textit{MCARI} & Vegetation index & $(B5 - B4) - 0.2(B5 - B3)(B5/B4)$ \\  
\textit{MCARI2} & Vegetation index & $1.5[2.5(B8 - B4) - 1.3(B8 - B3)] $ \\
&& $\qquad / \sqrt{(2 \times B8 + 1)^2 - (6 \times B8 - 5 \sqrt{B4}) - 0.5}$ \\ 
\textit{GNDVI} & Vegetation index & $(B8A - B3) / (B8A + B3)$ \\ 
\textit{MSAVI} & Vegetation index & $0.5[2 \times B8A + 1 - \sqrt{(2 \times B8A + 1)^2 - 8(B8A - B4)}]$ \\ 
\textit{CVI} & Vegetation index & $(B8A \times B5) / (B3^2)$ \\
\textit{SIPI3} & Vegetation index & $(B8 - B2) / (B8 - B4)$ \\
\textit{SR\_740\_705} & Vegetation index & $B6 / B5$ \\
\textit{REPLI} & Vegetation index & $700 + 40* (((B4 + B7)/2 - 5)/ (B6 - B5)) $\\ 
\textit{TCI} & Vegetation index & $1.2 * (B5 - B3) - 1.5 * (B4 - B3) * \sqrt{B5 / B4}$ \\
\textit{ARI} & Vegetation index & $(1/B3) - (1/B5)$ \\ 
\textit{CARI} & Vegetation index & $B5 / B4 \times \left| 670 (B5 - B3) / 150 + B4 + B3 - 550 (B5 - B3) / 150 \right| $ \\
&& $\qquad / \sqrt{(B5 - B3) / 150^2 + 1}$ \\
\textit{LCI} & Vegetation index & $(B8 - B5) / (B8 + B4)$ \\
\textit{MTVI1} & Vegetation index & $1.2[1.2(B8 - B3) - 2.5(B4 - B3)]$ \\
\textit{MTVI2} & Vegetation index & $1.5[1.2(B8 - B3) - 2.5(B4 - B3)] $ \\
&& $\qquad / \sqrt{(2 \times B8 + 1)^2 - (6 \times B8 - 5\sqrt{B4}) - 0.5}$ \\[4pt]

\hline
\end{tabular}
\end{table}

\section{Implementation, training and validation}\label{sec:implementation} 

All statistical models and computational procedures were implemented in R \citep{r_core_team_r_2021}.

\paragraph{Random Forests.}
For the stacked hurdle model dedicated to field-scale disease prediction, Random Forests were implemented using the \texttt{ranger} package \citep{wright_ranger_2017}. Hyperparameter calibration was performed with the \texttt{caret} package \citep{kuhn_building_2008} using a five-fold cross-validation procedure.

Covariate importance was computed for each learner to interpret the models and relate their driving factors to existing agronomic and remote-sensing knowledge. It was computed using the \texttt{ranger} package, based on impurity reduction \cite{scornet_trees_2023}. For any covariate $X_g$, $g\in\{1,\ldots,q\}$, its importance value is defined in Equation~\ref{eq:impurity}:
\begin{equation}
\label{eq:impurity}
\mathrm{Imp}(X_g)
= \sum_{v \in \mathcal{V}(X_g)} \Delta I_v ,
\end{equation}

where $\mathcal{V}(X_g)$ denotes the set of tree nodes at which covariate $X_g$ was used for splitting, and $\Delta I_v$ represents the decrease in node impurity resulting from the split at node $v$.

In classification trees, node impurity is measured using the Gini index:

\begin{equation}
\label{eq:gini}
I_v = 1 - \sum_{k=1}^{K} p_{vk}^2 ,
\end{equation}
where $p_{vk}$ is the proportion of samples of class $k$ at node $v$, and $K$ is the total number of classes. In a hurdle model, as it is the case here, the classification task is binary and $K = 2$.

For regression trees, node impurity corresponds to the variance of the response covariate:
\begin{equation}
\label{eq:var}
I_v = \frac{1}{n_v}\sum_{i \in v} \bigl( y_i - \bar{y}_v \bigr)^2 ,
\end{equation}
where $n_v$ is the number of observations in node $v$ and $\bar{y}_t$ is their mean.

To enable comparison of covariate importance across different Random Forest models, the importance values were normalized into percentages:

\begin{equation}
\label{eq:importance_percent}
\mathrm{RelImp}(X_g)
= 100 \times 
\frac{\mathrm{Imp}(X_g)}
     {\sum_{g'=1}^{q} \mathrm{Imp}(X_{g'})} .
\end{equation}

\paragraph{Stacked hurdle model meta features.} Meta-features that summarise local prediction from the initial learner at different spatial scales are based on a sequence of neighbourhood radii $0 = d_1 < d_2 < \cdots < d_L$.
In the application, we used the radii $d \in \{0,\ 10,\ 25,\ 50,\ 100,\ 200,\ 579\ \text{km}\}$ to provide a regular and interpretable set of spatial scales ranging from purely local information ($d=0$) to the maximum distance between two fields in the dataset (579 km).
This multi-scale design allows the model to capture both fine-scale and broad-scale spatial patterns in disease risk.

\paragraph{Disease progress curve.} The threshold value, $N^\star$, representing the minimum number of observations required to directly fit a field DPC to local severity values, was set to $3$ in order to reasonably estimate the two parameters $(r_j, t_j^{50})$ of the logistic disease progress curve. 

For the computation of DPC values, time was discretized on a regular grid $\tau_1 < \cdots < \tau_L$ over the epidemic season $\mathcal T$.
For each year, the first grid point $\tau_1$ was set to the date of the first available observation, and following grid points were defined with a $7$ days intervals. The grid was truncated at the last observation date of the year. This one-week temporal resolution was chosen as it was sufficient to capture the evolution of sugar beet yellows severity over time.

For sparsely observed locations, $u_j$ in $\mathcal{U}_{\mathrm{sparse}}$, the weight, $w_j$, is chosen so that all smoothed pseudo-observations together contribute as much as a single real observation:
\[ w_j = \frac{1}{N_j}. \]
where $N_j$ represents the number of local observations. This ensure that the influence of the smoothed information does not dominate the scarce real observations. Ideally, the weighting scheme should be calibrated jointly with $N^\star$ (which represents the upper limit of $N_j$ for $u_j$ in $\mathcal{U}_{\mathrm{sparse}}$) to prevent over-dilution of pseudo-observations.

\paragraph{Nadaraya-Watson spatial smoothers.}

The spatial bandwidth parameters of the Nadaraya-Watson kernel smoothers, respectively $h^{(0)}$ when smoothing estimated DPC values at temporally frequently observed locations, and $h$ when smoothing estimated DPC values at both frequently and sparsely observed locations, were selected by block cross-validation.
During the cross-validation process, each block corresponded to a French administrative department (see Figure 1B). For each block, field-level severities were recomputed by smoothing estimated severity values from all other blocks. The RMSE between DPC-based severities and cross-validated smoothed severities was then computed. The optimal bandwidth was defined as the value minimizing this cross-validated RMSE.

To generate continuous spatial predictions, severity values were then interpolated over a regular grid covering the study area. A grid with a spatial resolution of 2.5 km was created over all regions where at least one observation was available. Grid cells located farther than 50 km from the nearest monitored field were removed to avoid extrapolation artefacts. In addition, grid cells located more than 2.5 km away from any beet field (sugar, fodder, or table beet), as identified in the French Land Parcel Identification System (LPIS) for the corresponding year, were also excluded for more spatial relevance.

\paragraph{Model training \& coupling between observations and remote sensing data.} Field-scale stacked hurdle model  was trained to predict each year from 2019 to 2023, except 2022 for which data were not available. Each year was treated as independent. Thus, when the model was trained to predict a given target year, all other years were used for training, whether they occurred before or after the target year.

Because the epidemiological profiles differed across years, a variable proportion of data from the target year was also included in the training set. This proportion ranged from 0 to $0.9$, aiming at evaluating the ability of the model to adapt to year-specific conditions and to assess the potential complementarity between field observations and remote sensing data. The selection of data included in the dataset was based on stratified sampling. For each French administrative department, the same proportion of unique sites $u_j$ were randomly selected and added to the training dataset. The validation dataset then consisted of data from the target year that has not been used in training.

Epidemic dynamics were reconstructed using the spatio–temporal semi-parametric model only for years 2020 and 2023, during which a sufficient number of fields ($>50$) met the requirement of having at least three observations (i.e., belonged to $\mathcal{U}_{frequent}$). This was performed using the satellite-based predictions ($\hat{y_i}$) generated by the stacked hurdle model, with 20\% of the observations from the target year included in its training dataset. For both years, epidemic dynamics were also reconstructed from observed severity values ($y_i$), instead of predicted severity values ($\hat y_i$), as a benchmark.

\paragraph{Model performances.}

Model performances were evaluated using the Root Mean Square Error (RMSE) (Equation~\ref{eq:rmse_stacked_hurdle}). For the stacked hurdle model, predicted values ($\hat y_i$) were compared with ground-truth observations ($y_i$) on validation data:

\begin{equation}
\label{eq:rmse_stacked_hurdle}
\mathrm{RMSE} = \sqrt{\frac{1}{n} \sum_{i=1}^{n} (y_i - \hat{y}_i)^2} .
\end{equation}

For the spatio–temporal semi-parametric model, performance was assessed by RMSE (Eq. \ref{eq:rmse_spatio_temporal}) comparing epidemic dynamics reconstructed from satellite-derived predictions ($\hat{y}(s,t)$) and those obtained from direct field observations ($\hat{y}_{\text{benchmark}}(s,t)$), which served as a benchmark, at all spatial locations $s \in \mathcal{S}$ and time points $t \in \mathcal{T}$, where disease severity is estimated. This comparison quantifies how much information is preserved when epidemiological dynamic reconstruction relies on remote-sensing-derived estimates rather than direct observations.

\begin{equation}
\label{eq:rmse_spatio_temporal}
\mathrm{RMSE} = \sqrt{\frac{1}{|\mathcal{S}||\mathcal{T}|} \sum_{s \in \mathcal{S}; t \in \mathcal{T}} (\hat{y}_{\text{benchmark}}(s,t) - \hat{y}(s,t))^2} ,
\end{equation}
where $|\mathcal{S}|$ and $|\mathcal{T}|$ are the number of grid cells and time points covering $\mathcal{S}$ and $\mathcal{T}$, respectively.

\paragraph{Model uncertainty.} For the semi-parametric spatio temporal model, uncertainty was evaluated using a non-parametric cluster bootstrap, in which each spatial location $\{u_j\}_{j=1}^m$ is treated as a cluster.
$B$ denotes the number of bootstrap replicates and was set at $B = 100$. In each bootstrap iteration $b = 1,\dots,B$, the set of spatial units was resampled with replacement to form a new bootstrap dataset. For each bootstrap dataset, the semi-parametric spatio-temporal model applied to obtain a new estimate of the spatio-temporal severity surface. For any location $s$ and time $t$ with $(s,t)\in \mathcal{S}\times \mathcal{T}$, we denote the resulting bootstrap estimate by
\[\hat y^{(b)}(s,t), \qquad (s,t)\in \mathcal{S}\times \mathcal{T}.\]
The pointwise bootstrap standard error was then computed as the empirical standard deviation across the $B$ bootstrap replicates:

\begin{equation}
\label{eq:bootstrap_se}
\widehat{\mathrm{SE}}\!\left[\hat y(s,t)\right]
=
\operatorname{sd}\!\left(
  \hat y^{(1)}(s,t),\,
  \ldots,\,
  \hat y^{(B)}(s,t)
\right),
\qquad (s,t)\in\mathcal{S}\times\mathcal{T}.
\end{equation}

This approach provides a fully non-parametric estimate of the sampling variability of the reconstructed spatio-temporal surface.

\section{Yearly Stacked Hurdle model RMSE}

\begin{figure}
\centering
\includegraphics[width=15cm]{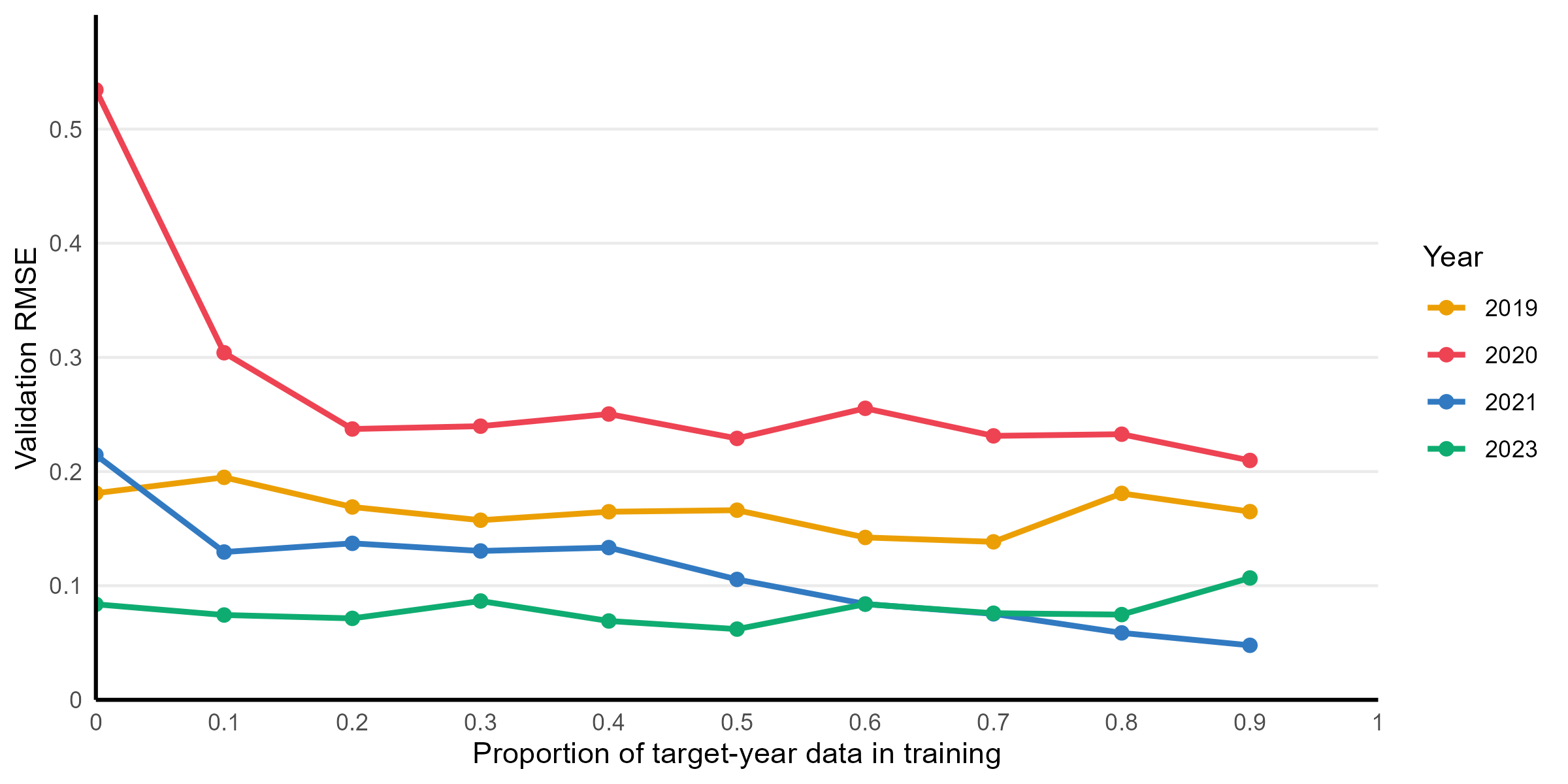}
\caption{\label{fig:rmse_prop_year} RMSE validation  versus the proportion of data from the predicted year included in the training dataset for all available years.}
\end{figure}

Figure \ref{fig:rmse_prop_year} shows the RMSE values of the stacked hurdle model as a function of the proportion of target-year data included in the training set for the four available years. For the year 2020, which was strongly affected by the severity of yellows disease across the entire territory, RMSE values are particularly high when no data from that year are included in the training set. This year is very difficult to predict based on other years, which have very different epidemic contexts. A similar, though less pronounced, pattern is observed for 2021. In contrast, for the years 2019 and 2023, which exhibit similar conditions, the proportion of target-year data included in training has little impact on RMSE values. Overall, this figure illustrates the difficulty of the model in producing reliable predictions when confronted with scenarios that are not represented in the training data.

\section{Classification \& probability of non-zero severity}

This figure is an alternative to Figure X, it focuses on the classification results from the final learner .

\begin{figure}
\centering
\includegraphics[width=15cm]{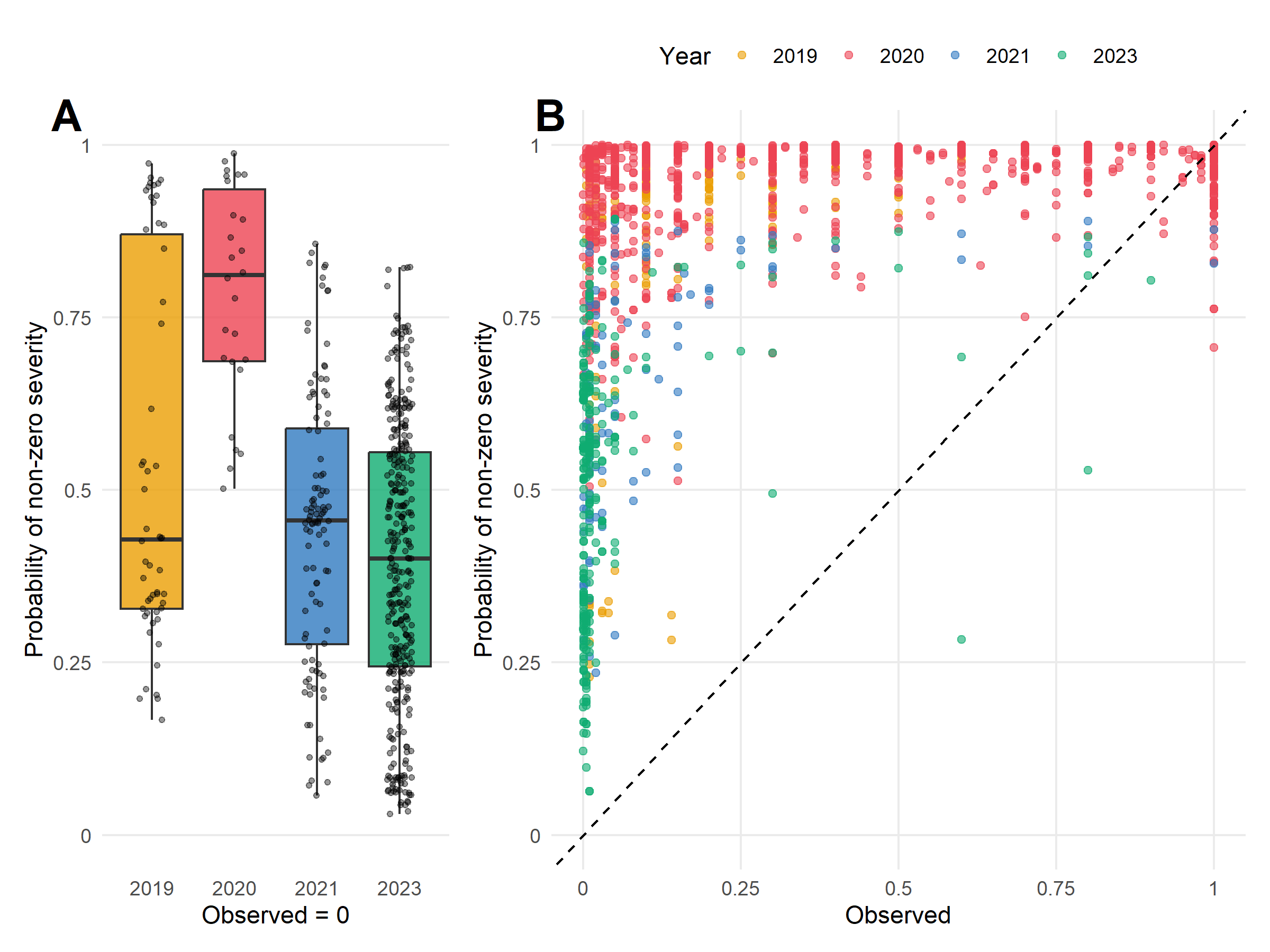}
\caption{\label{fig:obs_vs_proba} Comparison of observed values with probability of non-zero severity ($\hat p_i$) on validation data when 20\% of target year data is included in training. A: Probability where observed severity is zero, classified by year. B: Probability values where observed severity ranges between 0 and 1, 0 being excluded.}
\end{figure}

Figure \ref{fig:obs_vs_proba}A shows that, for fields with observed severity equal or close to zero, the predicted probability of non-zero severity is on average below 0.5 for the years 2019, 2021, and 2023. For 2020, this probability remains comparatively higher, reflecting the epidemiological context of that year were very few fields were unaffected. Overall, predicted probabilities for zero severity fields remain highly variable, indicating residual uncertainty in the classification stage.
For fields with moderate to high observed severity (Figure \ref{fig:obs_vs_proba}B), the classifier assigns consistently high probabilities of non-zero severity, with most values exceeding 0.75. This confirms the ability of the classification component to separate truly infected fields from those with no or negligible symptoms.

\section{Output maps}

\begin{figure}
\centering
\includegraphics[width=15cm]{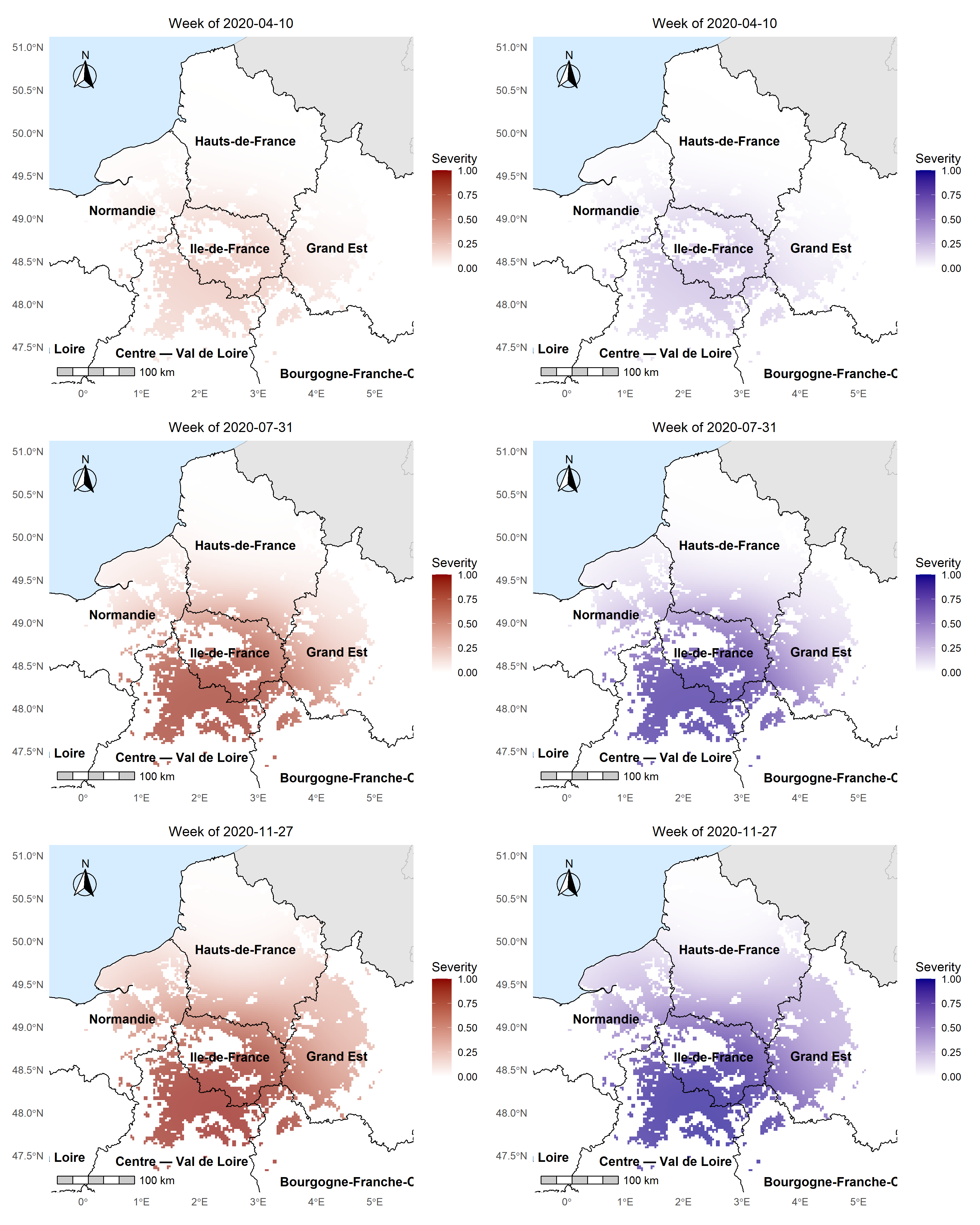}
\caption{2020 severity maps at the time of the first observations, mid-season, and at the end of the observation period. The maps on the left (red) are based on an epidemiological model calibrated with field observations, while the maps on the right (blue) are based on an epidemiological model calibrated with predictions from the final learner using satellite data.}
\label{fig:maps_2020}
\end{figure}

\begin{figure}
\centering
\includegraphics[width=15cm]{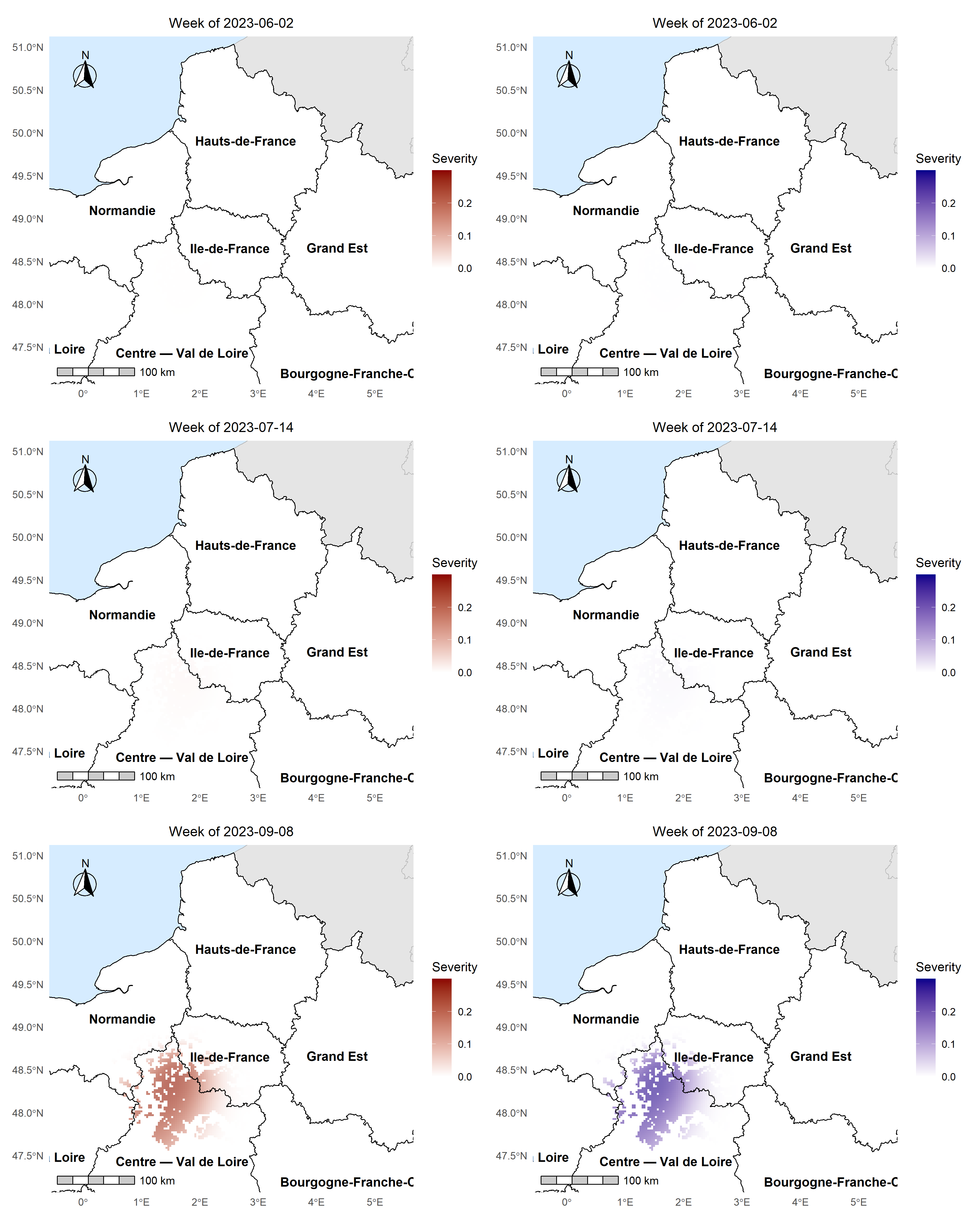}
\caption{\label{fig:maps_2023}2023 severity maps at the time of the first observations, mid-season, and at the end of the observation period. The maps on the left (red) are based on an epidemiological model calibrated with field observations, while the maps on the right (blue) are based on an epidemiological model calibrated with predictions from the final learner using satellite data.}
\end{figure}

Figures \ref{fig:maps_2020} and \ref{fig:maps_2023} illustrate the outputs of the spatio-temporal model at the production basin scale for the years 2020 and 2023, respectively.

\section{RMSE of the spatio-temporal semi-parametric model}

\begin{figure}
\centering
\includegraphics[width=15cm]{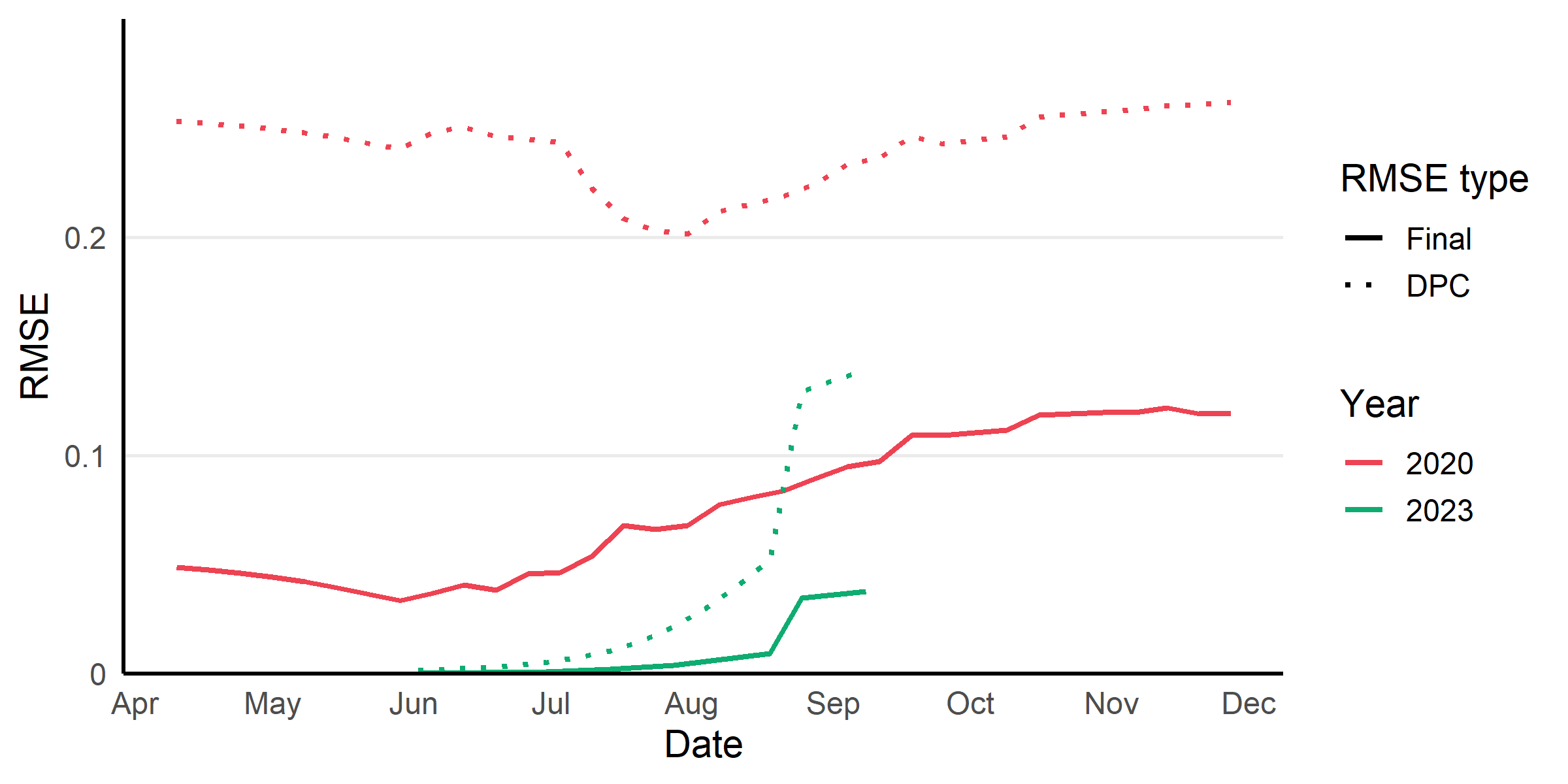}
\caption{\label{fig:smooth_dpc_rmse} Temporal evolution of the RMSE of the spatio-temporal semi-parametric model. Errors compare satellite-based predictions with field-based (benchmark) predictions of beet yellows severity for the 2020 and 2023 seasons. For each year, solid lines reprensent RMSE computed from the spatio-temporally smoothed severity estimates and dotted lines represent RMSE computed from the fitted disease-progress curves. Disease-progress curves are fitted over the full observation period for each season.}
\end{figure}

In 2020, a year marked by widespread beet yellows across much of the territory, RMSE values of the smoothed DPC predictions ranged between 0.03 and 0.09. These values are comparable to those obtained with the field-scale prediction model on which the spatio-temporal epidemiological model is based (Figure R2). 
In 2023, the RMSE values were much lower, especially early in the season when they were close to zero. As disease-progress curves start at zero severity, errors were generally smaller at the beginning of the season and increased slightly later on.
Across both years, RMSE values were consistently higher for the unsmoothed DPC predictions than for the spatio-temporally smoothed estimates. This highlights the robustness introduced by the smoothing procedure, which reduces local prediction noise and spatially isolated errors, enabling a more reliable reconstruction of the overall epidemic dynamics.

\section{Bootstrap maps}

\begin{figure}
\centering
\includegraphics[width=15cm]{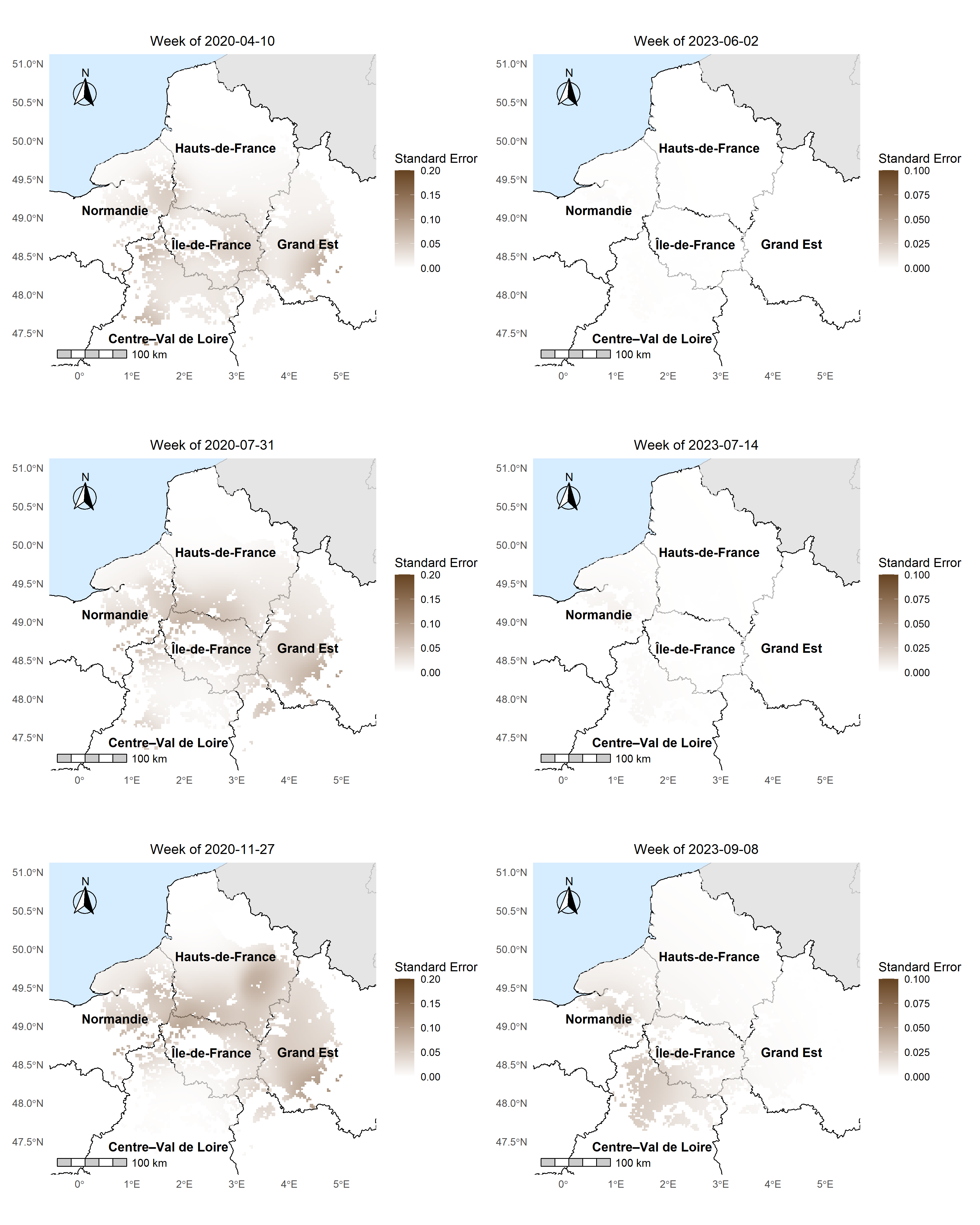}
\caption{\label{fig:maps_SE} Spatialized standard error of the semi parametric spatio-temporal model at the time of the first observations, mid-season, and at the end of the observation period for year 2020 (left) and 2023 (right). Standard errors are based on a bootstrap with $B=100$ repetitions.}
\end{figure}

Figure \ref{fig:maps_SE} illustrates the spatial distribution of standard errors obtained with the bootstrap of the spatio-temporal model.

\section{Interpretation of variable importance}

 From an agronomic perspective, variable importances (Figure 3) 
 are consistent with a strong spatial and temporal structuring as spatio-temporal metadata dominate relative importance across all four Random Forest models. From an agronomic perspective, the relative importances given to the different covariates are consistent with existing knowledge. Vegetation indices that contribute the most are those related to chlorophyll absorption ($MTCI$, $mNDblue$, $TCARI\_OSAVI$, $CARI$, $MCARI\_OSAVI$). This aligns with known physiological effects of beet yellows, which reduces leaf area development \citep{de_koeijer_effects_1999} and decreases leaf optical absorption \citep{clover_effects_1999}. The high importance of mNDblue is particularly noteworthy, as this index was specifically designed for monitoring chlorophyll activity in sugar beet under short-range proxy detection conditions \citep{jay_estimating_2017}. Other indices, though less influential, likely provide complementary cues: pigment-sensitive indices such as ARI or structure-related indices such as NDVI may help distinguish yellows from fungal diseases \citep{mahlein_spectral_2010}, while moisture-related indices such as NDWI \citep{gao_ndwinormalized_1996,bouasria_use_2021} may assist in separating virus symptoms from water-stress responses.
Raw Sentinel-2 bands show lower but notable importance, with red-edge bands (B5, B6, B7) likely helping the model to capture changes in leaf biochemistry, visible bands (B2, B3) contributing to evaluate chlorophyll content, and short-wave infrared bands (B11, B12) carrying information on canopy water content. 
For the final learners, the fact that the most relevant aggregation distances for meta-features range from 0 to 25 km is consistent with what is known about aphid flight patterns and the fact that they migrate preferentially over short distances \citep{loxdale_relative_1993}

\section{Alternative representation of the stacked hurdle model}

\begin{figure}
\centering
\includegraphics[width=15cm]{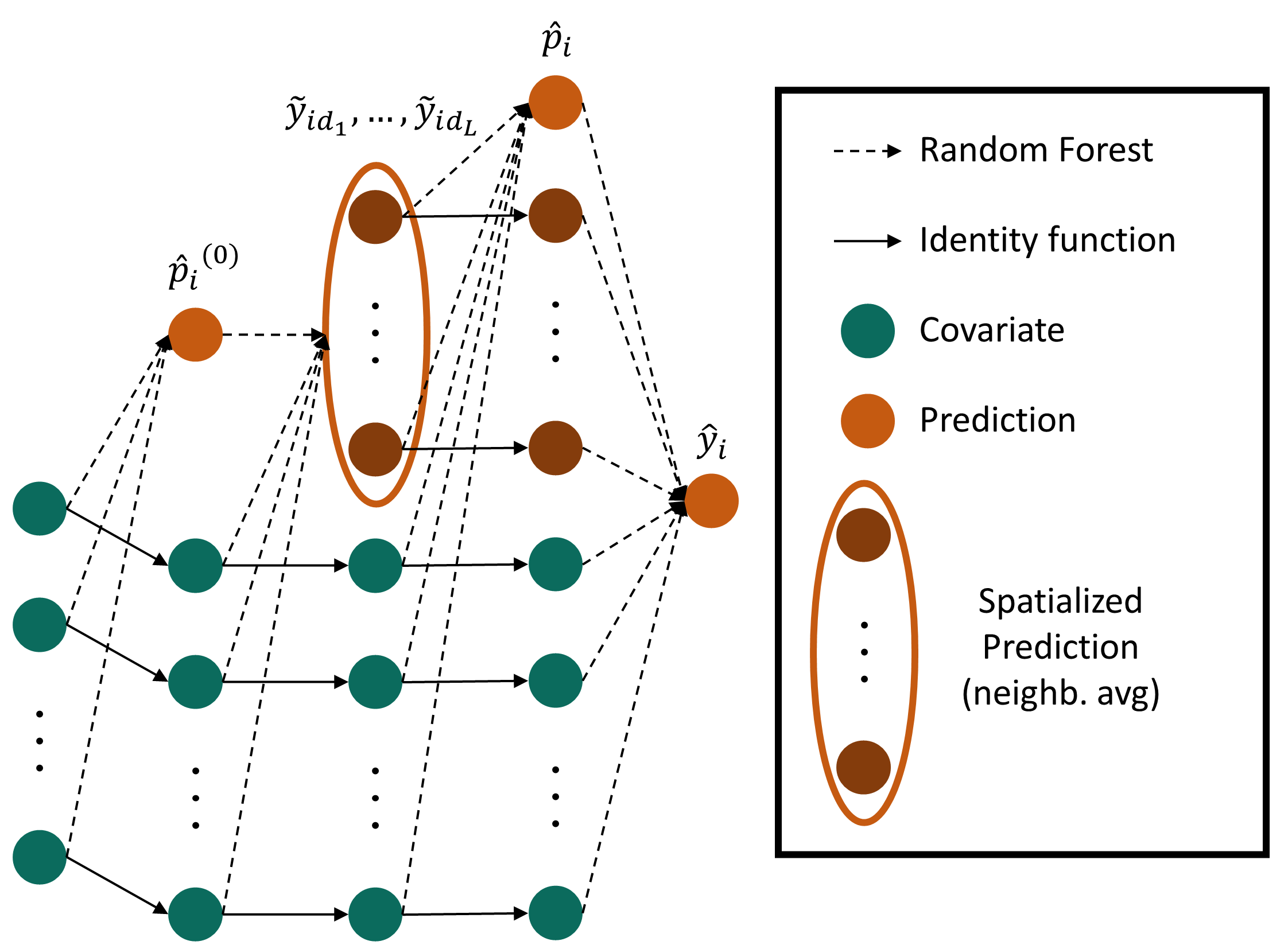}
\caption{\label{fig:neural_rep} Alternative representation of the stacked hurdle model as a pseudo-neural network.}
\end{figure}

The stacked hurdle model can be represented as a heterogeneous layered ensemble, sharing structural similarities with neural networks (Figure \ref{fig:neural_rep}). In the proposed model, several predictions are obtained in successive layers, each transforming the output of the previous one. The key distinction here lies in the nature of the transformations and their associated weights between layers. Unlike standard neural networks, where activation functions are generally the same in a given layer, the explicitly defined transformations are heterogeneous at the layer level in the proposed framework, with random forest for successive prediction and identity functions to propagate covariates across layers. Its structure can be adjusted by changing base learners, redefining meta-features, or adding layers to account for spatial structure, temporal dependence or other specificity without altering the overall structure of the model.

\bibliographystyle{plainnat}
\bibliography{sample}

\end{document}